\documentclass[fleqn,usenatbib]{mnras}

\usepackage{newtxtext,newtxmath}

\usepackage[T1]{fontenc}
\usepackage{ae,aecompl}

\usepackage{graphicx}	
\usepackage{amsmath}	
\usepackage{amssymb}	
\usepackage{bm}
\usepackage{times}
\usepackage{hyperref}
\usepackage[normalem]{ulem}
\usepackage{multirow}

\renewcommand{\jcap}{J. Cosmol. Astropart. Phys.}

\newcommand{\hMpc}{\ensuremath{ ~h^{-1} \rm{Mpc} }}	

\usepackage{color}
\definecolor{colorChanges}{rgb}{.8,.2,1.}
\definecolor{colorDeletes}{rgb}{1.,.2,.2}

\newcommand{\degSq}{\ensuremath{~\mathrm{deg}^2}}
\newcommand{\map}[1]{\ensuremath{#1 \times #1 \degSq}}
\renewcommand{\arcmin}{\ensuremath{~\rm{arcmin}}}

\defcitealias{Takahashi2017}{T17}
\defcitealias{Barreira2017}{B17}

\title[Testing gravity using WL voids]{Cosmological test of gravity using weak lensing voids}

\author[Davies et al.]{
Christopher T.~Davies$^{1}$\thanks{E-mail: christopher.t.davies@durham.ac.uk},
Marius Cautun$^{1,2}$
and Baojiu Li$^{1}$
\\
$^{1}$Institute for Computational Cosmology, Department of Physics, Durham University, South Road, Durham DH1 3LE, UK\\
$^2$ Leiden Observatory, Leiden University, PO Box 9513, NL-2300 RA Leiden, the Netherlands \\
}

\pubyear{2019}

\begin{document}
\label{firstpage}
\pagerange{\pageref{firstpage}--\pageref{lastpage}}
\maketitle

\begin{abstract}
Modifications to General Relativity (GR) often incorporate screening mechanisms in order to remain compatible with existing tests of gravity. The screening is less efficient in underdense regions, which suggests that cosmic voids can be a useful cosmological probe for constraining modified gravity models. In particular, weak lensing by voids has been proposed as a promising test of such theories.
Usually, voids are identified from galaxy distributions, making them biased tracers of the underlying matter field. An alternative approach is to study voids identified in weak lensing maps -- weak lensing voids -- which have been shown to better correspond to true underdense regions. 
In this paper, we study the ability of weak lensing voids to detect the signatures of modified gravity.
Focusing on the void abundance and weak lensing profiles, we find that both statistics are sensitive probes of gravity. These are quantified in terms of the signal-to-noise ratios (SNR) with which an LSST-like survey will be able to distinguish between different gravity models. We find that the tangential shear profiles of weak lensing voids are considerably better than galaxy voids at this, though voids have somewhat lower SNR than weak lensing peaks. The abundances of voids and peaks have respectively $\rm{SNR} = 50$ and $70$ for a popular class of modified gravity in an LSST-like survey.
\end{abstract}




\section{Introduction}

The large scale structure (LSS) of the Universe is the result of the anisotropic gravitational collapse \citep{Zeldovich1970} and it takes the form of an intricate pattern, the so-called cosmic web, which is made up of interconnecting knots, filaments, walls and voids \citep{Kirshner1981,Davis1985,Bond1996}. Knots correspond to the largest over-densities in the cosmic web, where matter flows through connecting walls and filaments \citep{Cautun2014,Haider2016}, while, in turn, the walls and filaments accrete mass from and enclose large under-dense voids, with the voids occupying most the volume of the Universe \citep[e.g.][]{Padilla2005,Platen2007,Cautun2013}. 

The formation of the cosmic web depends on the underlying cosmology, where various cosmological models result in different properties of the building blocks that make up the cosmic web. In particular, voids have been shown to be very useful cosmological probes, such as for measuring cosmological parameters \citep{Lavaux2012,Hamaus2015,Hamaus2016,Correa2019,Nadathur2019}, testing the nature of gravity \citep{Li2011,Clampitt2013,Barreira2015,Cai2015,Zivick2015,Achitouv2016,Cautun2018,Falck2018,Paillas2019}, the dark energy equation of state parameter \citep{Bos2012,Pisani2015,Demchenko2016} and the neutrino content \citep{Villaescusa-Navarro2013,Barreira2014,Massara2015,Banerjee2016,Kreisch2018}.

It is possible to observe the LSS through gravitational lensing, the bending of light due to intervening matter along the line of sight \citep{Bartelmann2001}. On cosmic scales, the projected LSS can be observed through weak gravitational lensing (WL) in the form of cosmic shear, that is small perturbations in the shapes of background galaxies \citep[e.g.][]{Bacon2000, Kaiser2000, VanWaerbeke2000, Wittman2000}. The properties of cosmic shear allows us to infer the cosmology of our universe \citep{Albrecht2006,LSST2012,Amendola2013,Weinberg2013}, through using statistics such as the shear-shear correlation function \citep[e.g.,][]{Schneider2002,Semboloni2006,Hoekstra2006,Fu2008,Heymans2012,Kilbinger2013,Hildebrandt2017} and the abundance of WL peaks \citep{Shan2012,Cardone2013,VanWaerbeke2013,Shan2014,X.Liu2015,J.Liu2016b,X.Liu2016,Higuchi2016,Shirasaki2017,Peel2018,Giocoli2018,Li2018,Davies2019}. 

One of the fundamental questions of cosmology concerns the cause of the accelerated expansion of the Universe, first detected by \citet{Riess1998} and \citet{Perlmutter1999}. Many possible explanations have been proposed \citep[e.g. see the recent review by][]{Caldwell2009}, but a very intriguing one concerns modifying gravity on large cosmological scales by including an extra scalar field, which mediates an additional, or fifth, force. However, GR has been shown to conform accurately with gravity tests in the Solar System \citep{Bertotti2003,Will2014}, and, since any modifications to GR must pass the same tests, it requires that the fifth force must be suppressed in our Solar System. One way to achieve this suppression is through screening mechanisms, where the effects of the fifth force only become important in under-dense regimes \citep{Brax2013}. One of such phenomenological models which contains a screened fifth force is the normal branch of the Dvali-Gabadadze-Porrati braneworld models (nDGP) \citep{Dvali2000}. In the nDGP model, the fifth force is suppressed through Vainshtein screening \citep{Vainshtein1972}, which is least effective far from massive objects, and so we expect that the greatest detectable signatures of the fifth force would be most apparent within voids. 

Given that WL maps correspond closely to the projected LSS, it is only natural to use them to identify structures such as high density peaks as well as low density regions. In this paper, we study the latter, i.e., {\it WL voids}, by employing the \citet{Davies2018} method of identifying voids in the WL convergence field. The objective is to study the potential of these WL voids to constrain modified gravity models. Our study was motivated by the results of \citet{Cautun2018} and \citet{Paillas2019} who found that voids identified in the galaxy distribution are emptier in modified gravity models compared to the standard cosmological model, $\Lambda$CDM, and that this signature can be measured in the tangential shear profile of voids. \citet{Davies2018} have shown that the tangential shear of WL voids is about $3$ times higher than that of galaxy voids and therefore WL voids represent a promising approach for testing MG models. We exemplify the constraining power of WL voids by studying the nDGP model above, and the results will have implications for upcoming surveys, such as LSST \citep{LSST2009} and Euclid \citep{Refregier2010}, which aim to provide high resolution WL maps over a large fraction of the sky. Studying WL voids represents a new approach of maximising the information that can be gained from such future data sets.

This paper is structured as follows: in Section \ref{sec:Theory} we discuss the relevant modified gravity and weak lensing theory, in Section \ref{sec:Numerical simulations} we present the data used in this study. We describe the prescription we follow to include galaxy shape noise (GSN) in our analysis in Section \ref{sec:GSN}. The void finder used in this work is described in Section \ref{sec:void finding algorithm}, followed by results for the WL peak abundance, void abundance, void convergence profile and void shear profile in $\Lambda$CDM and MG in Section \ref{sec:results}. We finally conclude in Section \ref{sect:disc}.

\section{Theory}
\label{sec:Theory}

In this section, for completeness, we very briefly describe the main points of the nDGP model and the weak lensing theory. 

\subsection{Modified gravity theory}

nDGP is a brane-world model in which the 4D spacetime (a brane) is embedded in a 5D spacetime called the bulk. Matter particles are confined to the brane, while gravitons can move through the extra dimensions of the bulk. A scalar field is introduced to represent the coordinate of the brane in the extra dimension, known as the brane-bending mode. The scalar field is a physical degree of freedom in the model which mediates a fifth force, and the strength of the fifth force is controlled by a crossover scale $r_c$, the scale at which the behaviour of gravitons changes through 4D or 5D.
The nDGP action for gravity is
\begin{equation}
    S = \int_{\rm{bulk}} d^{5}x \sqrt{-g^{(5)}} \frac{R^{(5)}}{16\pi G^{(5)}} + \int_{\rm{brane}} d^{4}x \sqrt{-g} \frac{R}{16\pi G} \, ,
\end{equation}
where $g$ is the determinant of the metric tensor, $R$ is the Ricci scalar, $G$ is the gravitational constant, all on the brane, and a superscript $^{(5)}$ denotes the 5D bulk counterparts to the the above 4D brane terms. The cross-over scale at which gravity transitions from 5D to 4D is related to $G$ and $G^{(5)}$ as
\begin{equation}
    r_c = \frac{1}{2} \frac{G^{(5)}}{G} \, ,
\end{equation}
and the modified Poisson equation receives an additional contribution from the scalar field $\psi$ in the form 
\begin{equation}
    \nabla^{2}\Psi = \nabla^{2}\Psi_{N} + \frac{1}{2}\nabla^2\psi \, ,
\end{equation}
where $\Psi_N$ is the standard Newtonian potential satisfying
\begin{equation}
    \nabla^2\Psi_{N} = 4\pi Ga^2\delta\rho \, ,
\end{equation}
$a$ is the scale factor, $\nabla$ is the spatial derivative and $\delta \rho = \rho - \overline{\rho}$, where $\rho$ is the matter density and $\overline{\rho}$ is the background matter density. 

The equation of motion of $\psi$ is given by 
\begin{equation}
    \nabla^{2}\psi + \frac{ r_{c}^{2} }{ 3\beta(a) a^2 } \bigg[( \nabla^{2} \psi)^{2} - \nabla^i\nabla^j\psi\nabla_i\nabla_j\psi \bigg] = \frac{8\pi G}{3\beta(a)}\delta \rho a^2 \, ,
    \label{eq:psi eom}
\end{equation}
where $i, j$ run through 1,2 and 3, and $\beta$, which dictates the strength of the fifth force, is a function of time given as
\begin{equation}
\begin{split}
    \beta(a) &= 1 + 2 H r_c \bigg( 1 + \frac{ \dot{H} }{3H^2} \bigg) \\
    &= 1 + \bigg[ \frac{\Omega_{m0}a^{-3} + \Omega_{\Lambda 0}}{ \Omega_{\rm{rc}} } \bigg]^{ \frac{1}{2} } - \frac{1}{2} \frac{ \Omega_{m0} a^{-3} }{ \sqrt{ \Omega_{m0} a^{-3} + \Omega_{\Lambda 0} } } \, ,
\end{split}
\end{equation}
where $H$ is the Hubble parameter, $H_0$ its present day value, $\dot{H}$ is its time derivative, $\Omega_{m0}$ is the present-day matter density parameter, $\Omega_{\Lambda 0}$ is the present-day vacuum energy density parameter and $\Omega_{\rm{rc}} = 1 / 4 H_0^2 r_c^2$. By linearising Eq.~(\ref{eq:psi eom}), the modified Poisson equation can be written as
\begin{equation}
\nabla^2\Psi = 4\pi G a^2 \bigg(1+\frac{1}{3\beta(a)} \bigg) \delta \rho \, . 
\end{equation}

Any modification to GR must pass the stringent Solar System tests of gravity, which means the fifth force must be well `screened' in environments like the Solar System, though it can still attain its full strength in under-dense regions. In the nDGP model, the fifth force is suppressed in over-dense regions through the Vainshtein mechanism in which, for an object in isolation, the radius within which screening is efficient is given by the Vainshtein radius $r_{V}$, 
\begin{equation}
    r_{V}^{3} = \frac{4GM}{9\beta^2H_0^2 \Omega_{rc}}.
\end{equation}
The fifth force becomes unscreened on scales $r \gtrsim r_V$.

\subsection{Weak lensing theory}
The convergence $\kappa$, for a single object along a line of sight, is linked to the lensing potential by
\begin{equation}
    \centering
    \kappa = \frac{1}{2} \nabla^2 \Psi_{\rm{2D}} \, ,
\end{equation}
where $\Psi_{\rm{2D}}$ is the lensing potential
\begin{equation}
    \Psi_{\rm{2D}}(\bm{\theta}) = \frac{D_{ls}}{D_l D_s} \frac{1}{c^2} \int_0^{z_s} \Phi_{\rm{len}}(D_l\bm{\theta},z) dz \, ,
\end{equation}
in which $\bm{\theta}$ is the sky coordinate of the lensed object, $D_{s}$, $D_l$ and $D_{ls}$ are respectively the angular diameter distances between the observer and source, observer and lens, and lens and source, $z_s$ the source redshift, $c$ the speed of light and $\Phi_{\rm{len}}$ is the gravitational potential that couples to photons (not matter) which determines photon geodesics. 

The distinction between different types of potentials is important for modified gravity, since in some models the fifth force acts only on the massive matter particles (e.g., the default nDGP model), while in other models the fifth force directly modifies the photon geodesics (e.g. our nDGPlens model). In this work we consider two MG models, nDGP and nDGPlens \citep{Barreira2017}, where the only difference between the two models is the form of $\Phi_{\rm{len}}$, which for nDGP is 
\begin{equation}
    \nabla^2\Phi_{\rm{len}}^{\rm{nDGP}} = \nabla^2\Phi_{\rm{len}}^{\rm{GR}} = 4 \pi G a^2 \delta \rho \, ,
    \label{eq: lcdm len pot}
\end{equation}
and which for the so-called {\it nDGPlens} model is given by
\begin{equation}
    \nabla^2\Phi_{\rm{len}}^{\rm{nDGPlens}} = \nabla^2\Phi_{\rm{len}}^{\rm{GR}} + \frac{1}{2} \nabla^2 \psi = 4 \pi G a^2 \delta \rho + \frac{1}{2} \nabla^2 \psi.
    \label{eq: nDGPlens lens.pot.}
\end{equation}
These imply that in nDGPlens, the lensing of photons receives an extra contribution from the scalar field, when compared to $\Lambda$CDM and nDGP. nDGPlens is created by us to illustrate the behaviour of a MG model where photon geodesics are modified as well, and such models do exist in the literature, such as the cubic Galileon model studied in \citep{Barreira2015}\footnote{Note that in the nDGP model considered here the accelerated expansion is driven by an additional dark energy species, which we tune to ensure that the background expansion history is identical to that of $\Lambda$CDM, while in the Galileon model self-acceleration by the scalar field can be achieved and the expansion history is generally different from $\Lambda$CDM. The nDGPlens model is chosen to have the same expansion history as nDGP, and so it is different from cubic Galileon, and is only used as a toy model to single out the effect of modified photon geodesics.}.

The previous equations apply to the lensing induced by a single lens, however for cosmic shear it is important to consider lensing contributions from all matter along the line of sight. So $\kappa$ can be written more generally as

\begin{equation}
    \kappa(\bm{\theta}) = \int_0^{z_s} W(z) \delta \rho(D_l(z)\bm{\theta},z) dz \, ,
    \label{general lensing with kernel}
\end{equation}
for a lensing potential given by Eq. \eqref{eq: lcdm len pot}, i.e., one that is the same as the GR case. Where $W(z)$ is the lensing kernel that includes the redshift distribution of the multiple lenses, given by
\begin{equation}
    W(z) = \frac{3 H_0^2 \Omega_{m0}}{2 c} \frac{1+ z}{H(z)} \chi(z) \int_{z}^{z_s} \frac{dn}{dz_s} dz_s \frac{\chi(z_s) - \chi(z)}{\chi(z_s)} \, ,
\end{equation}
$\chi$ denotes the comoving distance, and $\frac{dn}{dz_s}$ is the redshift distribution of sources. If, however, the lensing potential is modified as in Eq. \eqref{eq: nDGPlens lens.pot.}, $\kappa$ (in the linear regime) becomes
\begin{equation}
    \kappa(\bm{\theta}) = \int_0^{z_s} W(z) \bigg(1 + \frac{1}{3 \beta(\alpha)}\bigg) \delta \rho(D_l(z)\bm{\theta},z) dz \, ,
    \label{general lensing with kernel mod}
\end{equation}
which indicates that the convergence values will be rescaled by a constant (in space) across a WL map. Due to the Vainshtein screening, however, the MG effect on $\kappa$ is more complicated and can only be accurately predicted through simulations. 

\section{Weak lensing maps}
\label{sec:Numerical simulations}

All the convergence maps used in this work cover a \map{10} sky area and have a resolution of $2048^2$ pixels per map and a source redshift of $z_s = 1$. Throughout this work, we will make use of WL maps generated from two sets of simulations. The first data set we use are three  WL maps from \cite{Barreira2017} (hereafter \citetalias{Barreira2017}) with $\Lambda$CDM, nDGP and nDGPlens cosmologies respectively, generated from the modified N-body code {\sc ecosmog} \citep{Li2012,Li2013} and ray tracing performed with {\sc ray-ramses} \citep{Barreira2016}, a code that implements the on-the-fly ray tracing algorithm proposed by \citet{White:1999xa,Li-ray-tracing}. These maps are used to predict the differences of several lensing and void observables between the different gravity models. Secondly, in order to generate covariance matrices and error bars used in the SNR analysis in Section \ref{sec:results}, we use the all-sky $\Lambda$CDM WL maps from \cite{Takahashi2017} (hereafter \citetalias{Takahashi2017}) which we split into 184 non-overlapping \map{10} maps following the method presented in \cite{Davies2019}.

\subsection{Numerical simulations}

The lightcone geometery used to generate the WL maps from \citetalias{Barreira2017} consists of seven tiled dark-matter-only simulation boxes, of which the first five (the ones closest to the observer) have a box size of $L = 300 \hMpc$ and the remaining two boxes, in order of increasing distance from the observer, have sizes $L = 350 \hMpc$ and $L = 450 \hMpc$. Each of the seven N-body simulations were run using a particle number of $N_p = 512^3$. The cosmological parameters used for \citetalias{Barreira2017} were the fractional baryon density $\Omega_b = 0.049$, fractional dark matter density $\Omega_{\rm{dm}} = 0.267$, dimensionless Hubble rate $h = H_0 / 100$ km s$^{-1}$ Mpc$^{-1} = 0.6711$, primordial scalar spectral index $n_s = 0.9624$ and root-mean-squared (rms) density fluctuation smoothed over $8\hMpc$ $\sigma_8 = 0.8344$. For a more detailed description of the simulation procedure used to generate the \citetalias{Barreira2017} maps we refer the reader to \cite{Barreira2017,Barreira2016}.

For the WL maps from \citetalias{Takahashi2017}, a series of dark matter-only simulation boxes with comoving sizes $L, 2L, 3L, ..., 14L$ where $L = 450 \hMpc$ are produced. Each of these simulation boxes are duplicated eight times and then nested around the observer, such that the larger boxes enclose and overlap with the smaller boxes (see Figure 1 of \cite{Takahashi2017} for an illustration). Ray tracing was performed with the algorithm from \cite{Hamana2015}, on the mass distribution from the nested simulation boxes projected onto spherical shells with a thickness of $150 \hMpc$. The simulation used a partical number of $2048^3$ and cosmological parameters $\Omega_b = 0.046$, $\Omega_{\rm{dm}} = 0.233$, $\sigma_8 = 0.820$, $n_s = 0.97$ and $h = 0.7$. For a more detailed description see \cite{Takahashi2017}.

\subsection{Galaxy shape noise}
\label{sec:GSN}

Weak lensing maps obtained from observational data require measurements of redshifts and shapes for a large number of background galaxies. Intervening cosmic structure acts as a lens for the source galaxies and induces small correlations in galaxy shapes across the sky, from which the cosmic shear signal can be extracted. However the amplitude of this correlation is small, and is largely dominated by the random orientation of galaxies, which is referred to as galaxy shape noise (GSN). 

Here, we are interested in characterising how well the difference between the $\Lambda$CDM and MG models, can be measured observationally so we add GSN to all of the convergence maps used in this study. We smooth the convergence maps with a compensated Gaussian filter $U$ \citep{Hamana2012}, which satisfies $\int^{\theta_o} U(\theta) \theta d\theta = 0$, where
\begin{equation}
    \label{eq: filter}
    U = \frac{1}{\pi \theta_s^2} e^{-\theta^2 / \theta_s^2} - \frac{1}{\pi \theta_o^2} (1 - e^{-\theta_o^2 / \theta_G^2})
\end{equation}
for $\theta < \theta_o$, and $U = 0$ otherwise, with $\theta_s = 2.5\arcmin$, and $\theta_o = 15\arcmin$. This choice of filter allows us to account for the mass sheet degeneracy \citep{Schneider1996}, and removes the long wavelength modes. For the peak and void abundances discussed in Sections \ref{WL peak abundance} and \ref{sec: void abundance} we also show results for the $\Lambda$CDM case using Gaussian smoothing (with 
$\theta_s = 1.5\arcmin$). We make this comparison since both the compensated Gaussian filter \citep[e.g.][]{Hamana2012,Shirasaki2018} and the Gaussian filter \citep[e.g.][]{J.Liu2015} are common choices for weak lensing studies, and we include SNR values for both filters for all statistics measured throughout this work. This comparison allows us to demonstrate the impact the choice of filter can have on measurements made in WL maps, beyond variations in the smoothing scale.

Our prescription for including GSN is based on \cite{VanWaerbeke2000b}, where we add random values to each pixel taken from a Gaussian distribution with standard deviation, $\sigma_{\rm{pix}}$, given by
\begin{equation}
    \sigma_{\rm{pix}}^2 = \frac{\sigma_{\rm{int}}^2}{2 \theta_{\rm{pix}} n_{\rm{gal}} }
    \;,
\end{equation}
where $\sigma_{\rm{int}}$ is the dispersion of source galaxy intrinsic ellipticity, $\theta_{\rm{pix}}$ is the angular width the pixels in the WL maps, and $n_{\rm{gal}}$ is the number density of source galaxies. In order to make a forecast for LSST, we use $\sigma_{\rm{int}} = 0.4$ and $n_{\rm{gal}} = 40 $ arcmin$^{-2}$ \citep{LSST2009}.

For consistent definitions between the different cosmological models, we define the amplitude $\nu$ of a $\kappa$ pixel as
\begin{equation}
\nu = \frac{\kappa}{\sigma_{\rm{GSN}}},
\label{eq: nu_def}
\end{equation}
where $\sigma_{\rm{GSN}} = 0.007$ is the standard deviation of the GSN map (smoothed with the compensated filter) that is added to the data.

\section{Void finding algorithm}\label{sec:void finding algorithm}
\begin{figure*}
    \centering
    \includegraphics[width=2\columnwidth]{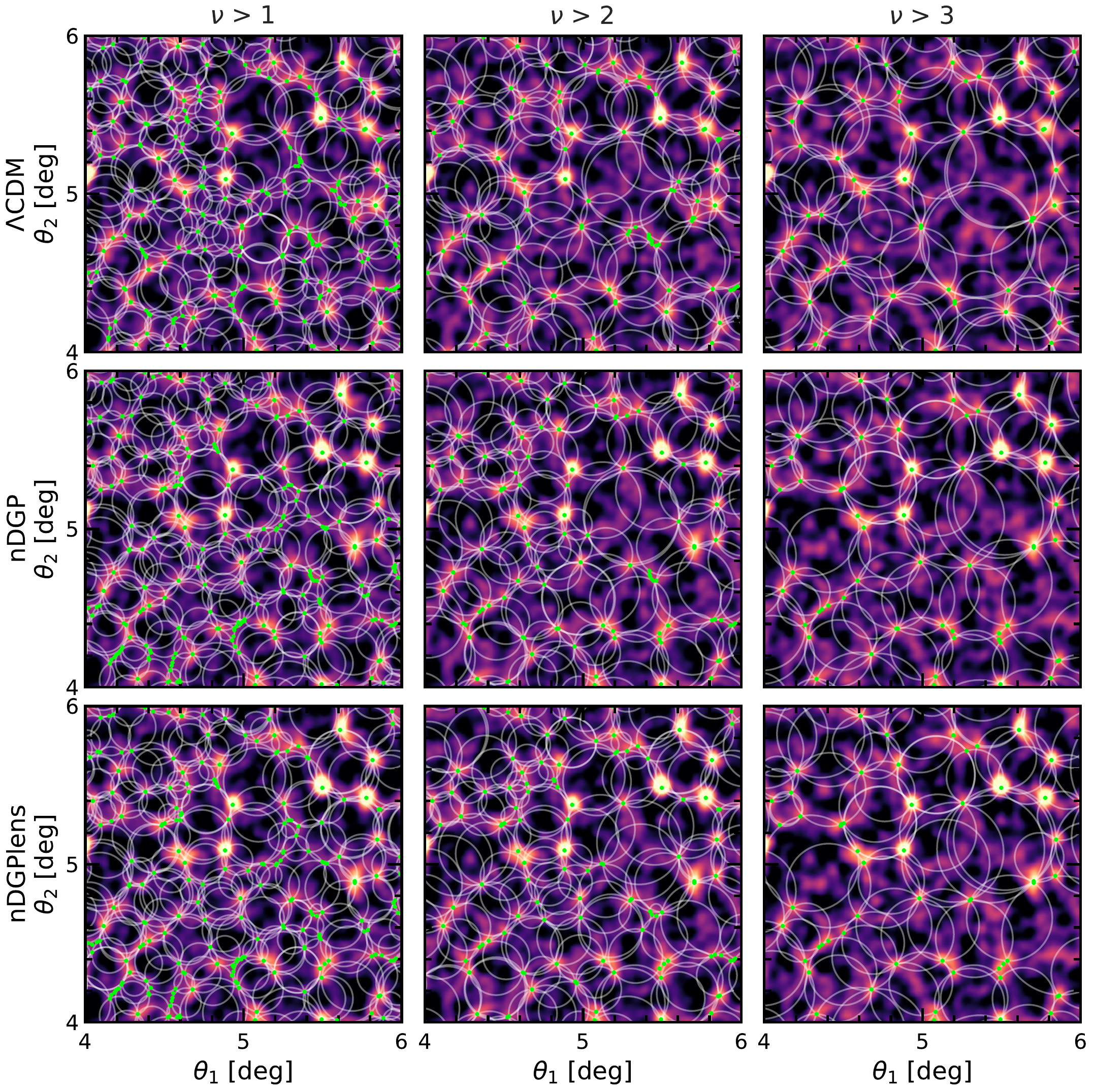}
    \vskip -.2cm
    \caption{ Illustration of the convergence field, and the peak and void catalogues in the $\Lambda$CDM (top row), nDGP (middle row) and nDGPlens (bottom row) models studied here. The $\kappa$ values are shown by the background colours, with bright colours corresponding to high $\kappa$ values and dark colours to low $\kappa$ values. 
    The axes $\theta_1$ and $\theta_2$ are two orthogonal angular coordinates, and only the central regions of the $\kappa$ maps are shown to avoid overcrowding the visualisation. 
    The WL peaks identified in these $\kappa$ maps are indicated by the green points, with the three columns corresponding to peaks of different heights: $\nu > 1 $ (left column), $\nu > 2$ (middle column) and $\nu > 3$ (right column).
    The white circles show the size and distribution of voids for each of the three peak catalogues.
    }
    \label{fig:void_vis}
\end{figure*}

In this paper we apply the tunnel algorithm of \cite{Cautun2018} to find voids. This is a 2D void finding algorithm which identifies voids based on an input tracer catalogue. This algorithm first constructs a Delaunay tessellation with the tracers as its vertices of the cells, and then voids are identified as the circumcircles of every Delaunay triangle, which is, by definition, empty of tracers. A void's centre corresponds to the centre of its respective Delaunay circumcircle and the void size, $R_{\rm{v}}$, is given by the radius of the respective circumcircle.

To apply the tunnel algorithm to WL maps, we use WL peaks as the input tracer catalogue. This produces 2D voids found in the WL convergence maps that, by definition, are devoid of WL peaks, with the closest peaks being found on the boundaries of the voids. To deal with the boundaries of the map, for the void abundance we remove any voids whose distance from the boundary is smaller than their radius, for the convergence and tangential shear ($\gamma_t$) profile plots we remove voids whose centres are within $2R_{\rm{v}}$ from the map boundary.

Furthermore, in order to increase the number of voids, which is necessary because of the small area of our WL maps, we consider all possible voids, including neighbouring ones which have a large degree of overlap (i.e., we do not exclude small voids which overlap with larger ones). The convariance matrix calculation, which is based on a much larger number of $\Lambda$CDM maps, ensures that the duplicate information is counted accordingly.

To identify WL peaks, we first smooth the convergence maps with a compensated Gaussian filter with smoothing scale $\theta_s = 2.5\arcmin$. From the smoothed WL maps, we identify WL peaks as pixels whose convergence values are larger than those of their eight neighbours. The peak catalogue used for the tunnel algorithm is created using information about the position and height of the WL peaks in the WL maps. For a given WL map and its associated peak population, we obtain three peak catalogues by selecting the peaks according to their height. The catalogues are comprised of peaks higher than a given $\nu$ threshold, with $\nu > 1, 2$ and $3$. For each WL map, we generate void catalogues from each of the three peak catalogues.

A visualisation of the tunnels identified in the WL maps studied here is shown in Fig. \ref{fig:void_vis}, where each row corresponds to one of the three models studied here. The columns correspond to peaks of different heights, and the associated void catalogues, with $\nu > 1,2$ and $3$ from left to right. It is evident from this figure that the $\nu >1$ peak catalogues produce the most voids, while the $\nu > 3$ catalogues produce more large voids. This means that the different void catalogues should respond to the large scale modes of the $\kappa$ maps differently, and so it is possible that the tightest constraints may be achieved through a combination of all three void catalogues, however due to the limited sample, this remains to be tested. The differences between $\Lambda$CDM and MG in Fig.~\ref{fig:void_vis}, can be studied quantitatively using peak and void abundances as well as void WL profiles, which is the subject of the next section.

\section{Results}
\label{sec:results}

In this section we discuss the properties of voids identified in WL maps, and present signal-to-noise-ratios (SNR) for the peak abundance, void abundance and tangential shear profiles as measures of the ability to distinguish between MG and GR. 

We define the SNR for a given statistic $S$ as
\begin{equation}
    {\rm{SNR}}^2 \equiv \sum_{i,j} \delta S(i) \,\, {\rm{cov}}^{-1}(i,j) \,\, \delta S(j)
    \label{eq:SNR} \;,
\end{equation}
where $\delta S = S_{\rm{MG}} - S_{\rm{GR}}$ is the difference in that statistic between MG and standard GR, $\rm{cov}^{-1}$ is the inverse of the covariance matrix for the statistic $S$ and $i$ and $j$ indicate the bin numbers that are summed. We multiply the $\rm{cov}^{-1}$ term by the Anderson-Hartlap factor $\alpha$ \citep{Anderson2003, Hartlap2007} in order to compensate for the bias present when inverting a noisy covariance matrix. The Anderson-Hartlap factor is given by
\begin{equation}
    \alpha = \frac{ N - N_{\rm{bin}} - 2 }{N - 1}
    \; ,
\end{equation}
where $N = 184$ is the number of realisations (WL maps) used to estimate the covariance matrix, and $N_{\rm{bin}}$ is the number of bins. The covariance matrices used for SNR measurements, using the compensated Gaussian filter, are shown and discussed in Appendix \ref{sec: cov_mat}. The covariance matrices used for the Gaussian filter are qualitatively similar to the compensated Gaussian case, and so we do not include them for brevity. The SNR values that we present in this work are forecast for LSST so we rescale the SNR values calculated from the $A = 100$ deg$^2$ maps by $\sqrt{ A_{ \rm{LSST} } / {A} } = 13.4$, assuming LSST will achieve a sky coverage of $A_{ \rm{LSST}} = 18000$ deg$^{-2}$.


\begin{table}
    \centering
    \caption{Forecasted SNR with which an LSST-like survey could discriminate between the two MG models studied here and $\Lambda$CDM. We show SNR values for WL peak and void abundance, as well as for the void tangential shear profile. For voids, we consider three different catalogues that were identified using the distribution of WL peaks with heights, $\nu > 1, 2$ and $3$.
    }
    \begin{tabular}{ ccccc } 
        \hline\hline
        $\nu$ range & \multicolumn{2}{c}{Compensated Gaussian} & \multicolumn{2}{c}{Gaussian}\\ 
         & nDGP & nDGPlens & nDGP & nDGPlens \\
        \hline \\[-.25cm]
        \multicolumn{5}{c}{\bf peak abundance, $n(> \nu)$ -- Fig. \ref{fig:peak_abundance} } \\[.1cm]
        $1\leq\nu\leq5$ & 71 & 146 & 110 & 195 \\
        \hline \\[-.25cm]
        \multicolumn{5}{c}{\bf void abundance, $n(> R_{\rm{v}})$ -- Fig. \ref{fig:voles_size_func} } \\[.1cm]

        $\nu > 1$ & 44 & 58 & 29 & 32 \\
        $\nu > 2$ & 68 & 49 & 29 & 44 \\
        $\nu > 3$ & 50 & 42 & 49 & 53 \\
        \hline \\[-.25cm]
        \multicolumn{5}{c}{\bf void tangential shear, $\gamma_t(r)$ -- Fig. \ref{fig:gamma_profile} } \\[.1cm]

        $\nu > 1$ & 46 & 80 & 51 & 68\\
        $\nu > 2$ & 50 & 68 & 41 & 59 \\
        $\nu > 3$ & 51 & 64 & 35 & 48\\
        \hline\hline
    \end{tabular}
    \label{tab:snr}
\end{table}

\subsection{WL peak abundance} \label{WL peak abundance}

\begin{figure}
    \centering
    \includegraphics[width=\columnwidth]{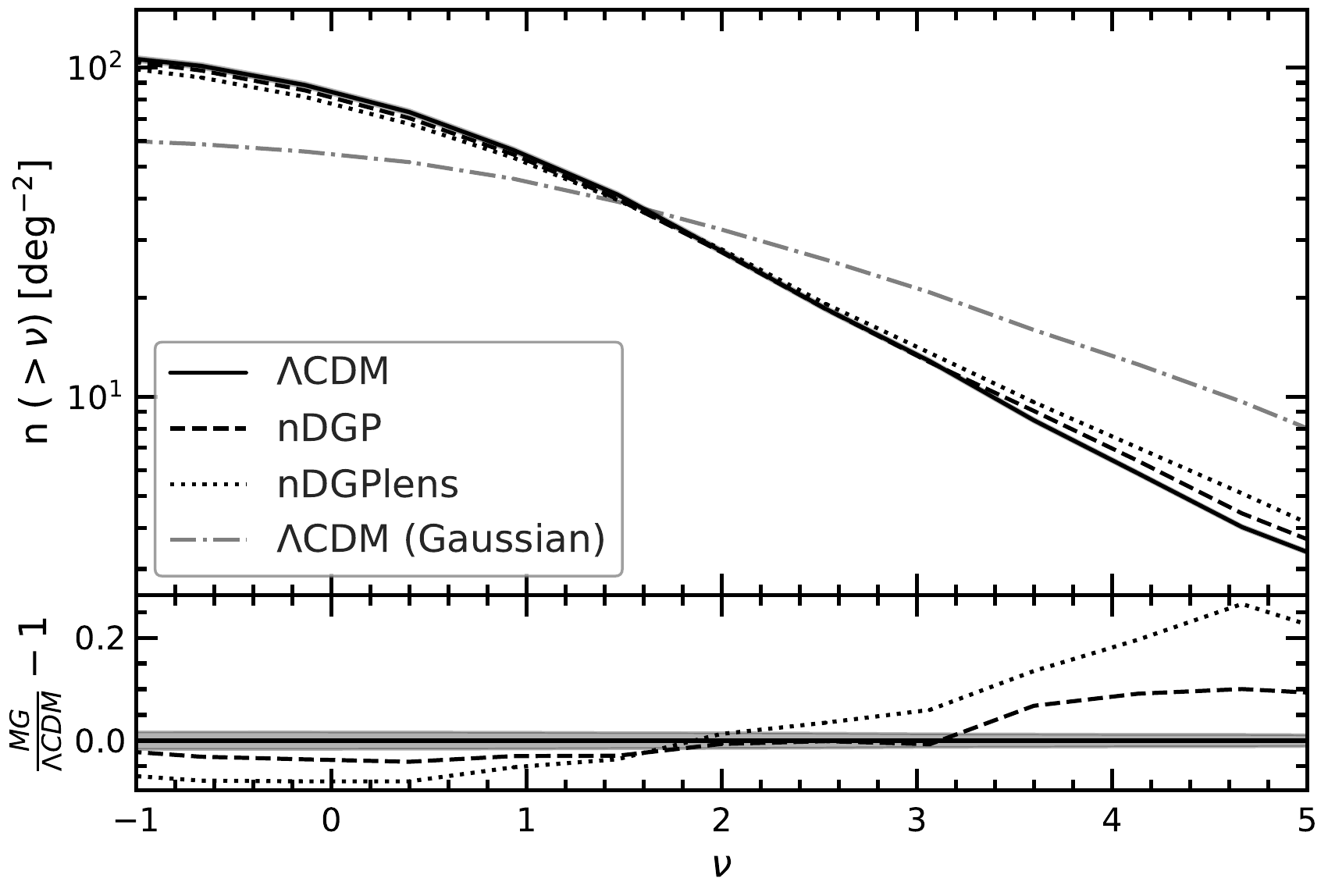}
    \vskip -.2cm
    \caption{
    Top panel: the WL peak abundance for $\Lambda$CDM (solid), nDGP (dashed) and nDGPlens (dotted) plotted as a function of peak height, $\nu$. The shaded regions indicate $1\sigma$ uncertainties for the $\Lambda$CDM result expected for an LSST-like survey.
    The WL peak abundance for $\Lambda$CDM smoothed with a Gaussian filter is shown by the grey dot-dashed line 
    Bottom panel: the relative difference of the peak abundance between the MG models and $\Lambda$CDM. Only the error bars for the $\Lambda$CDM curve have been plotted for clarity.
    }
    \label{fig:peak_abundance}
\end{figure}

Whilst the primary purpose of the WL peaks in this work is to be used as tracers for void identification, it is also interesting to consider how their abundance is affected by the MG models.

In the top panel of Figure \ref{fig:peak_abundance}, we show the number density of WL peaks as a function of peak height, $\nu$, for $\Lambda$CDM, nDGP and nDGPlens. The bottom panel shows the difference between the MG models and the fiducial $\Lambda$CDM one. The shaded regions in the figure (which are very small) indicate the uncertainties with which the $\Lambda$CDM peak abundance will be measured by LSST and are obtained from the peak abundance covariance matrix calculated using the \citetalias{Takahashi2017} maps. In the top panel, we can see that the modified gravity models produce slightly fewer small peaks with $\nu < 2$. For $\nu > 2$, nDGP and nDGPlens produce a higher number of large peaks than $\Lambda$CDM at fixed $\nu$, which is a consequence of the enhanced structure formation present in these MG models. This difference is present even for peaks with $\nu > 3$, which typically correspond to massive haloes, whose growth in the nDGP model is significantly enhanced. In each instance, the nDGPlens models shows the largest deviation from $\Lambda$CDM. Whilst the matter distribution is the same in nDGP and nDGPlens, the extra contribution to the lensing potential from the scalar field in nDGPlens allows for further modifications to the final WL maps, which boosts the amplitude of the peaks and thus results in more peaks for a fixed $\nu$ value.

The WL peak abundance for the $\Lambda$CDM map smoothed with a Gaussian filter ($\theta_s = 1.5$) is shown by the grey dot-dashed line in the top panel of Figure \ref{fig:peak_abundance}. The exact value of $\theta_s$ impacts the overall amplitude of the peak abundance, but the shape of the curve remains largely unaffected by changes in $\theta_s$, and so we choose a smoothing scale for the Gaussian filter that gives roughly the same number of peaks with $\nu>2$ as the compensated Gaussian case. The grey dot-dashed line illustrates that the Gaussian filter produces a shallower curve than the compensated Gaussian filter. This is possibly because more large peaks are identified with the Gaussian filter since the peaks receive contributions to their height from the large scale modes, which in the case of over-densities, will boost their height. These large scale modes are removed with the compensated Gaussian, which produces fewer large peaks. The affect this has on the void abundance is discussed in section \ref{sec: void abundance}

The differences in peak abundance between various MG models and the fiducial cosmology can be used as a cosmological test. For example, \citet{X.Liu2016} have shown that the WL peak abundance in the Canada-France-Hawaii-Telescope Lensing Survey \citep{Erben2013} can be used to make tight constraints on the parameters of $f(R)$ gravity. Motivated by this, the first row in Table \ref{tab:snr} shows the SNR with which LSST data for the peak abundance can distinguish between the MG models studied in this paper and $\Lambda$CDM. We calculated the SNR using all peaks with $ 1 \leq \nu \leq 5$, since peaks $\nu < 1$ are most likely to be contaminated by GSN, and peaks with $\nu > 5$ will be subject to stronger influence by sample variance due to the small sizes of the available WL maps. The SNR values for the Gaussian smoothing case are larger since this filter identifies more large peaks, which correspond to large haloes that are more likely to receive a boost in their growth due to MG. The model differences are qualitatively similar between the Gaussian and compensated Gaussian cases. In the next subsection, we study the extent to which the void abundance can also be used as a cosmological test.

\subsection{Void size function}\label{sec: void abundance}

\begin{figure}
    \centering
    \includegraphics[width=\columnwidth]{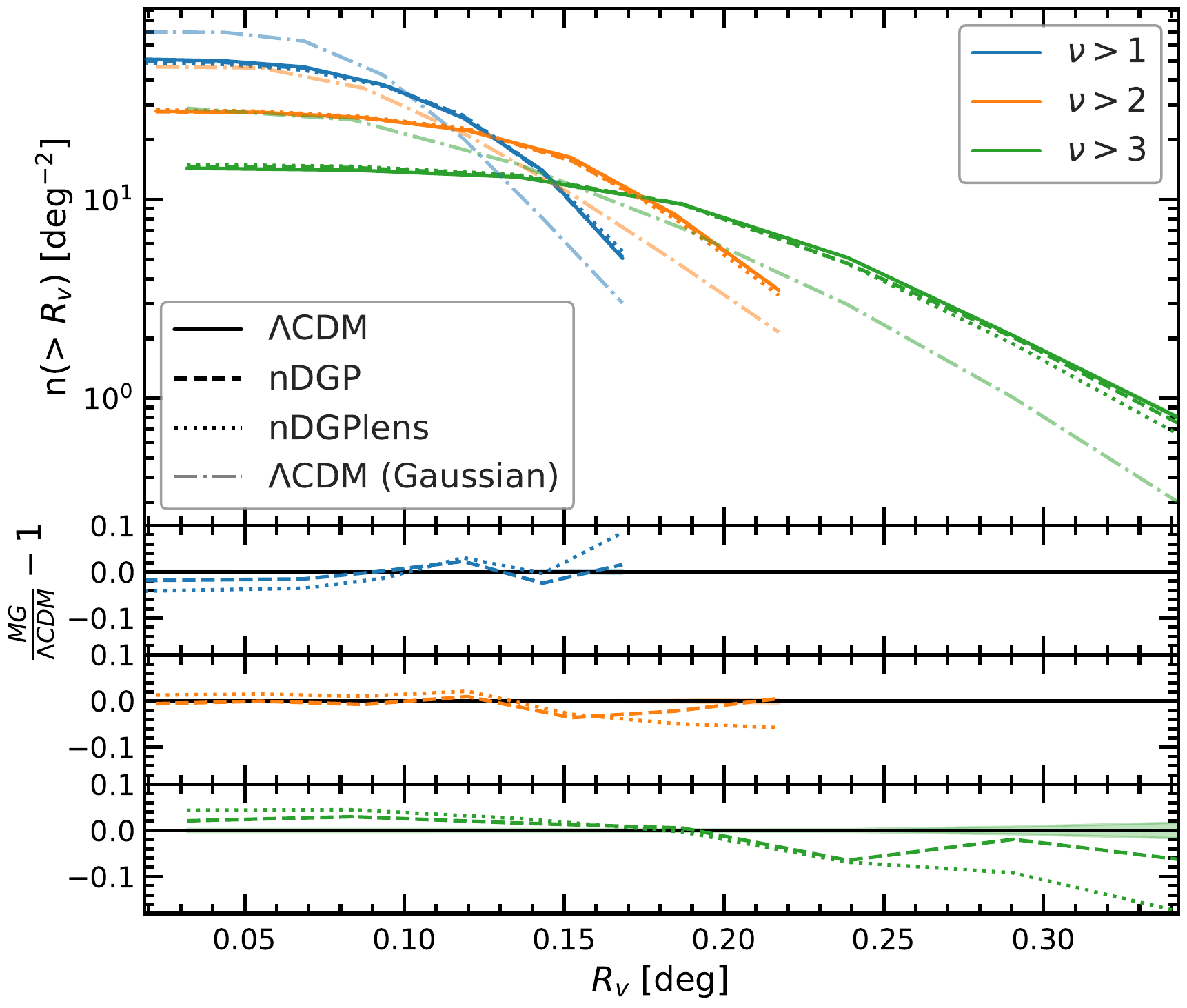}
    \vskip -.2cm
    \caption{Top panel: the void abundance as a function of void radius. The coloured curves correspond to voids identified in the WL peak distribution with heights: $\nu > 1$ (blue), $\nu > 2$ (orange), $\nu > 3$ (green). The void abundance for $\Lambda$CDM is shown by the solid line, nDGP is shown by the dashed line and nDGPlens is shown by the dotted line. The shaded region around the $\Lambda$CDM curve indicates $1\sigma$ error bars expected for a LSST-like WL survey (the error bar is roughly the same size as the thickness of the curves).
    Bottom panel: the relative difference between the void abundances in MG models and the fiducial $\Lambda$CDM cosmology for the three void catalogues shown in the top panel.
    }
    \label{fig:voles_size_func}
\end{figure}

Fig.~\ref{fig:voles_size_func} shows the distribution of void sizes for voids identified from the three WL peak catalogues that we study here, with $\nu > 1, 2$ and $3$. The error bars are calculated from the void abundance covariance matrix obtained using the 184 \citetalias{Takahashi2017} maps and are scaled up to the area of the LSST survey. As we have already seen from Fig.~\ref{fig:void_vis}, the smallest voids are generated by the $\nu > 1$ WL peak catalogue, which also produces the most voids. As the $\nu$ threshold increases, the typical void size increases, however there are fewer voids overall. For $\nu > 1$ and $2$, the total peak abundance is similar between $\Lambda$CDM and MG, which yields similar void abundances for all models in the $\nu >1$ and $2$ void catalogues. For $\nu >3$, there are more peaks for MG than for $\Lambda$CDM, which manifests itself as creating more small voids and fewer large voids in MG than in $\Lambda$CDM. This is a consequence of the larger number of peaks in the MG models that end up splitting large voids into several smaller ones. We note that the (very few) largest voids in each catalogue are not plotted in Fig. \ref{fig:voles_size_func}, and are also left out of the SNR calculation, since the differences between the models appear to be dominated by sample variance. This allows us to give more conservative estimates of the SNR for the void abundance, which are less affected by noise.

The SNRs with which the void abundance measurements in an LSST-like survey can distinguish between $\Lambda$CDM and the considered MG models are shown in middle rows in Table \ref{tab:snr}. In all cases, the void abundances produce lower SNR values than the peak abundance. The $\nu > 1$ catalogue produces the largest SNR values for nDGPlens, and the $\nu > 2$ catalogue gives the highest SNR for nDGP. Although the model differences between $\Lambda$CDM and nDGPlens are larger, the SNR values are not consistently higher than for nDGP. This is due to the relation between various entries of the covariance matrix that arises from the fact that fewer large voids imply more smaller voids, and thus the void abundance signature of nDGPlens is more similar to the trends expected from $\Lambda$CDM sample variance than for the nDGP case.

The faded dot-dashed lines show the void abundance for the Gaussian smoothing case. For the $\nu > 1$ catalogue, there is a larger total number of voids, and fewer large voids, than in the compensated Gaussian case.  
This is unexpected since the abundance of peaks with $\nu>1$ is lower for the Gaussian filter case than for the compensated Gaussian one, and typically having fewer peaks implies fewer voids too. However, we find that most of the extra peaks in the compensated Gaussian case are very highly clustered along the void boundaries for the $\nu > 1$ catalogue, and these peaks do not contribute additional voids to the catalogue despite contributing additional peaks. These highly clustered peaks on the void boundary for the $\nu > 1$ catalogue can be seen in the left column of Figure \ref{fig:void_vis}. The void abundance SNR values for the Gaussian smoothing case are on average lower than those for the compensated Gaussian one.

The voids are generated from the spatial distribution of the WL peaks, and hence depend on the clustering of these peaks. One way to measure this is through the N-point correlation functions of the peak catalogues. Therefore, an alternative way to exploiting the void abundance is to study the N-point correlation functions of the WL peaks, or the cross correlations between the void centres and the WL peaks. However, we find that within the limited statistics of our small maps, the 2-point correlation functions of WL peaks do not  show any significant differences between $\Lambda$CDM and the MG models. 

\subsection{Convergence profiles}
\begin{figure}
    \centering
    \includegraphics[width=\columnwidth]{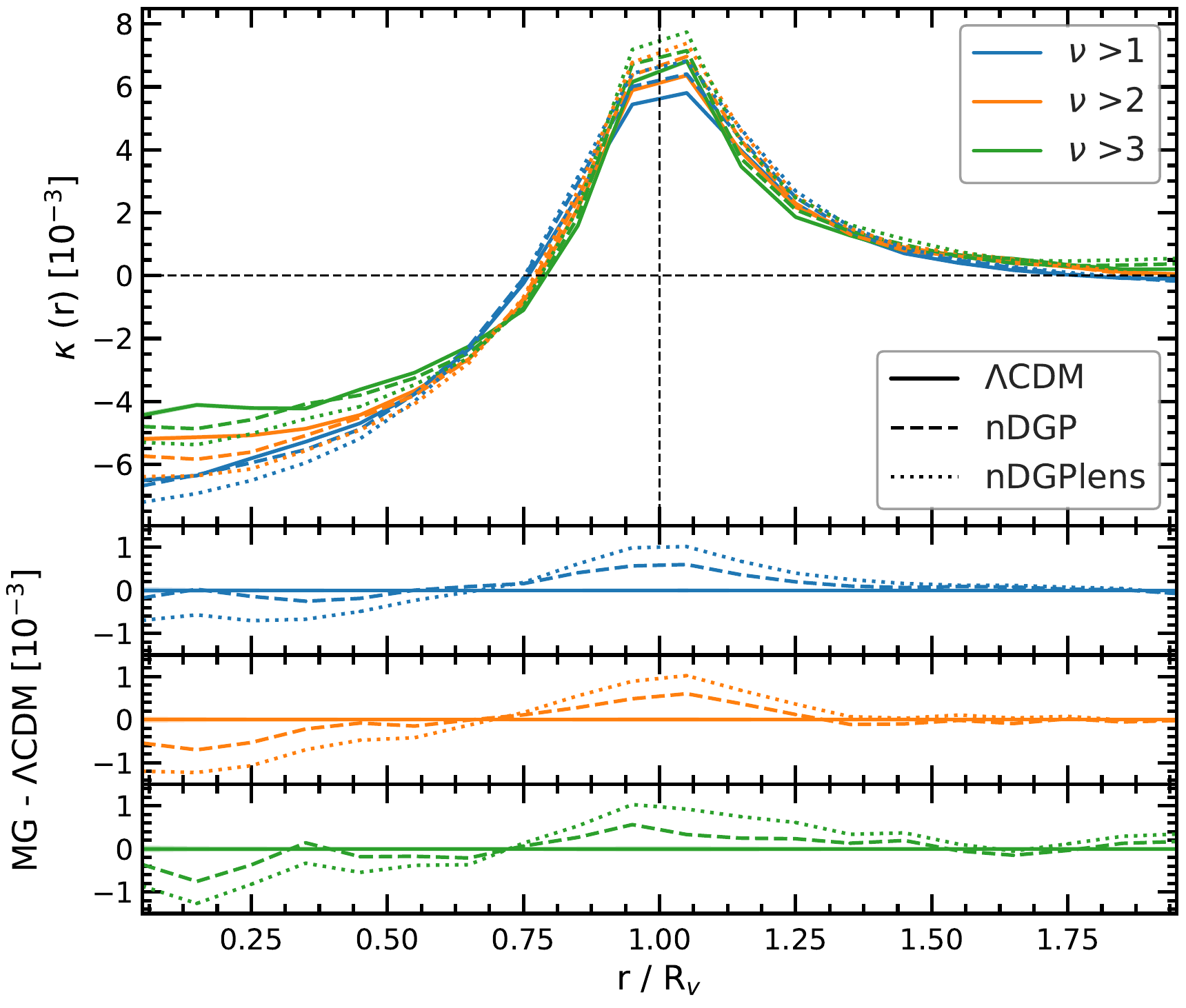}
    \vskip -.2cm
    \caption{Top panel: The stacked radial convergence profiles of the voids shown in Fig.~\ref{fig:void_vis} (excluding those within 2$R_{\rm{v}}$ from the map boundary). The three coloured curves correspond to voids identified from the three WL peak catalogues with  $\nu >1$ (blue), $\nu >2$ (orange) and $\nu > 3$ (green) respectively. The $\Lambda$CDM model is shown by the solid line, nDGP by the dashed line and nDGPlens by the dotted line. The shaded regions around the $\Lambda$CDM results show the $1\sigma$ error bars for an LSST-like survey. 
    Bottom panel: The relative difference of the $\kappa$ profiles between the MG models and $\Lambda$CDM.}
    \label{fig:kappa_profile}
\end{figure}

In the MG models studied here, the fifth force enhances structure formation, which results in more underdense voids than in the fiducial GR case \citep[e.g.][]{Falck2018}, with the excess matter that was evacuated from voids being deposited in the walls and filaments of the cosmic web that surround the voids \citep{Cautun2016,Paillas2019}. These differences in the clustering of matter manifest themselves in both the distribution of voids (as we seen in the previous section) and in the density profiles of voids. The $\kappa$ values in a WL map correspond to the projected matter density weighted by the lensing kernel and thus the differences in the matter content of voids are likely to be manifested also in the void $\kappa$ profiles. In this section, we study the mean convergence profiles of our three void catalogues and compare these profiles between different cosmological models.

We calculate the average $\kappa$ profile of voids by stacking all the voids in a given catalogue. Since the void size can vary by a factor of several between the largest and the smallest voids, we stack the voids in terms of the rescaled radial distance from the void centre, $r/R_{\rm{v}}$, i.e., we express the distance in units of the void radius.
Note that while the WL voids are identified in the smoothed WL maps, for calculating the $\kappa$ profiles we use the unsmoothed converge map. Using instead the smoothed $\kappa$ map results in shallower void profiles because the smoothing ``redistributes" the high $\kappa$ values found at a void's edge over the entire area of the void.

Figure \ref{fig:kappa_profile} shows the $\kappa$ profiles for each void catalogue in each of the cosmological models studied here. Similarly as before, the barely visible shaded regions correspond to error bars for an LSST-like survey and were obtained form the covariance matrix calculated from the 184 \citetalias{Takahashi2017} maps. The overall shape of the profile shows that the voids identified in the WL maps correspond to projections of under-dense structures since for $r < 0.75 R_{\rm{v}}$ all $\kappa$ profiles have negative convergence values. The $\kappa$ profile peaks at $r = R_{\rm{v}}$ for the void catalogues with the height of the maximum increasing as the $\nu$ threshold of the WL peak catalogues increases. The depth of the under-densities decreases as the $\nu$ threshold increases. This is because the void catalogues with a larger $\nu$ threshold contain larger voids, and since larger voids cover a larger number of pixels  of the WL map (whose mean $\kappa$ value is $0$), their profiles in general tend to $0$. It can also be seen that the regions outside of the void boundary remain over-dense at least up to a radial distance, $r=2R_{\rm{v}}$ for the $\nu > 3$ catalogue, with the smaller $\nu$ catalogues returning to $0$ at lower $r/R_{\rm{v}}$.

In general, we find that the void interiors ($r<0.75 R_{\rm{v}}$) in MG models have slightly lower $\kappa$ values than the corresponding points in $\Lambda$CDM (this effect is most readily visible in the $\nu > 1$ and $2$ catalogues, while for the other void catalogue the signal is slightly more noisy, in good agreement with previous studies \citep[e.g.,][]{Falck2018,Paillas2019}. Once the $\kappa$ profiles of voids become over-dense at $r / R_{\rm{v}} = 0.75$ the MG profiles become more over-dense than the $\Lambda$CDM ones out to $r = 2 R_{\rm{v}}$. The maximum difference in the $\kappa$ profiles between each model can be found at the void radius $r = R_{\rm{v}}$. 

\subsection{Tangential shear profiles}
\begin{figure}
    \centering
    \includegraphics[width=\columnwidth]{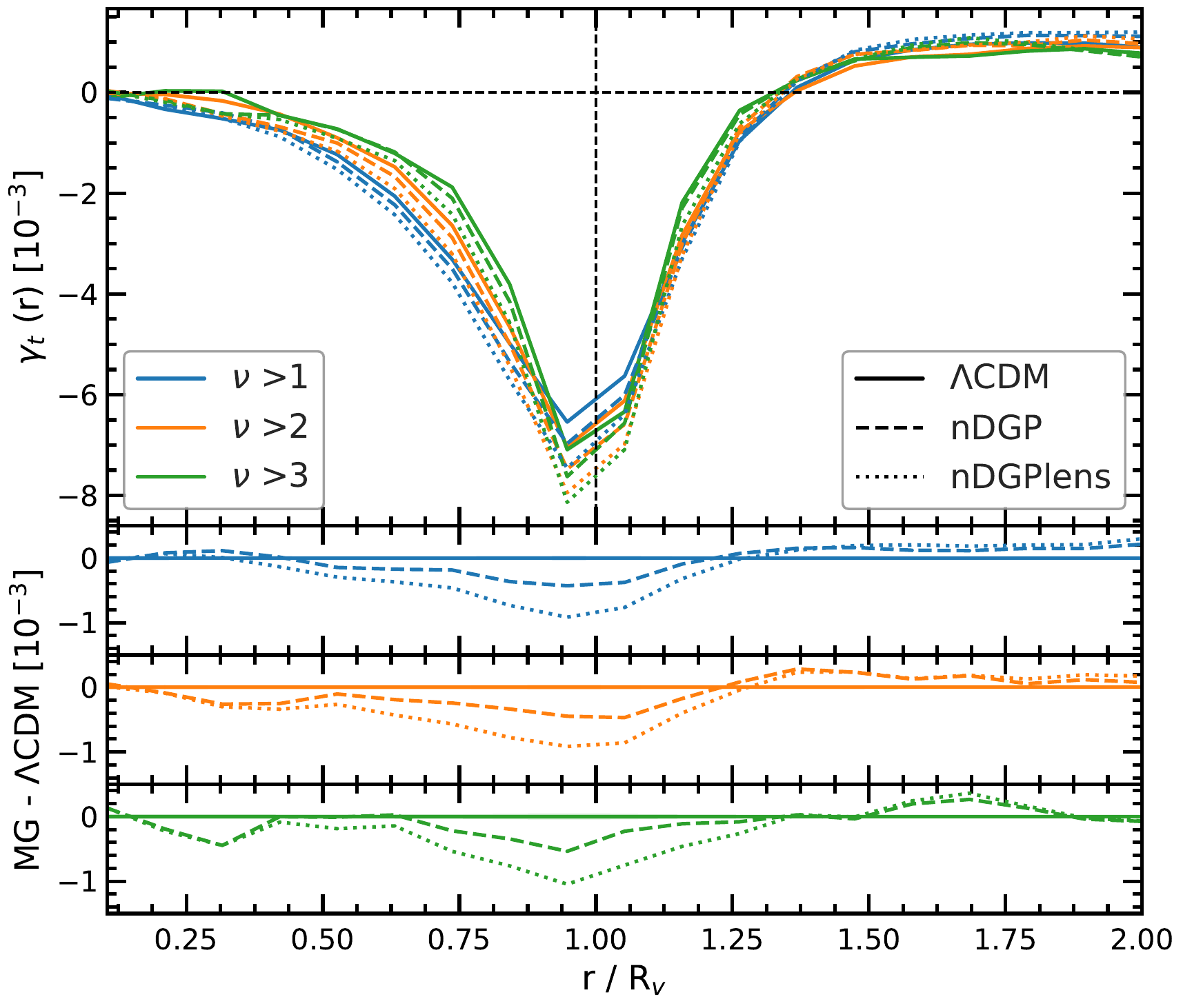}
    \vskip -.2cm
    \caption{The same as Fig. \ref{fig:kappa_profile} but for void tangential shear profile $\gamma_t$.
    }
    \label{fig:gamma_profile}
\end{figure}

Next we calculate the tangential shear profile $\gamma_t (r)$, for the different void catalogues. The tangential shear profile can be related to the convergence profile through
\begin{equation}
    \gamma_t(r) = \overline{\kappa}(< r) - \kappa (r)
    \label{eq:gamma_t} \;,
\end{equation}
where
\begin{equation}
    \overline{\kappa}(< r) = \frac{1}{\pi r^2}\int_0^{r} 2 \pi r' \kappa(r') dr',
    \;
\end{equation}
is the mean enclosed convergence within radius $r$. Whilst the convergence profiles of voids allow for a simple physical interpretation of their mass content, where positive and negative $\kappa$ correspond to projected over-dense and under-dense regions, it is the shear which can be measured directly in observations. Therefore, to more easily compare with observations, we also study the void tangential shear profiles. 

Fig.~\ref{fig:gamma_profile} shows the tangential shear profiles for the three void catalogues studied here. The typical shear value is negative indicating that voids act as concave lenses that bend light outwards from the void centres. It can be seen that the tangential shear peaks at $r = R_{\rm{v}}$ and the amplitude of this peak is largest for the $\nu >3$ catalogue.

Voids in MG models have larger tangential shear profiles, and the difference is the largest for the nDGPLens model, in which the fifth force both enhances structure formation and also directly affects the photon geodesics. To quantify the potential of void $\gamma_t$ profiles as a cosmological test, we summarise in the bottom three rows of Table \ref{tab:snr} the SNRs with which an LSST-like survey can distinguish the MG models studied here from $\Lambda$CDM.

Table \ref{tab:snr} shows that the SNR values of the $\gamma_t$ profiles in the nDGP model are roughly the same for all three void catalogues, although the $\nu > 3$ catalogue has a slightly larger SNR. For the nDGPlens model the SNR vary more between the various void catalogues, being the largest for $\nu > 1$ case. Overall the SNR values are highest for the nDGPlens model. Again, we find that on average the compensated Gaussian filter can discriminate MG model better than the Gaussian filter.

It was found in \cite{Paillas2019} that, with the same void finder and the same nDGP variant (N1), galaxy voids give an SNR value of 20, whereas here we find that voids identified directly in weak lensing maps produce SNR values up to 51. This shows that voids identified in weak lensing maps are ideal objects for studying the tangential shear profile. This is further motivation for the use of voids identified in weak lensing maps as complimentary statistics to the WL peak abundance, two-point correlation function and power spectrum. 

\section{Discussion and conclusions}
\label{sect:disc}

We have investigated the potential of voids identified in WL maps to distinguish between $\Lambda$CDM and a popular class of MG models, nDGP and its variant nDGPlens. For this, we smooth the WL maps with a 2.5 arcmin compensated Gaussian filter, before we identify WL peaks and use them as tracers for the tunnel void finding algorithm. We have then done a forecast for LSST, in which GSN is properly included and found that the WL void statistics, such as abundances and tangential shear profiles, are different in MG models compared to $\Lambda$CDM and can distinguish between $\Lambda$CDM and MG up to an SNR of about 80. The SNR values from $\gamma_t(r)$ for voids identified in WL maps are over two times larger than those of galaxy voids, making a strong case for the use of voids identified in WL maps as a complimentary probe of the LSS, and as a test of gravity. 

Throughout the paper, for the void abundance and the tangential shear profiles, we have used a range of $\nu$ thresholds in order to generate multiple void catalogues. However, from Table \ref{tab:snr}, there is no clear systematic trend which would indicate the best choice of $\nu$ threshold. Given the large range of void sizes in Fig.~\ref{fig:voles_size_func}, it is possible that each catalogue will respond to the small and large scale modes of the WL maps differently, and so there is potential for the multiple catalogues to provide complimentary information to each other. Due to the small map sizes available for this study, we leave a further investigation of this to future work. 

We find that the peak abundance gives larger SNR values than either the WL void abundance or the tangential shear profiles. This indicates that it is likely that the peak abundance will be able to provide tighter constraints on MG. However, the extent to which voids identified in weak lensing maps provide complimentary constraining power to the peak abundance remains to be studied, as the two statistics may respond differently to the changes in structure formation induced by MG, or have different degeneracy directions with other cosmological parameters. Additionally, it is possible that the results for the voids in this study are not fully converged due to the limited sample size: because the voids are physically larger than the peaks, it is possible that the voids require a larger sample area than the peaks before the measured SNR values are robust to changes in map area.

In order to reliably constrain MG with future surveys, further systematics must be taken into account. This includes the effect of baryons on simulated WL convergence maps, since we have used dark matter only simulations in this work. The full extent to which baryons alter WL statistics from dark matter only simulations depends on the sub grid model used. \cite{Yang2013} found that there is a significant amplitude increase in the WL power spectrum, and that low amplitude WL peaks remain unaffected by baryons, whilst the number of large peaks is increased by the inclusion of baryons. \cite{Weiss2019} found that in order for WL statistics from dark matter only and hydro simulations to agree, very large smoothing scales must be used ($8-16$ arcmins), which is partly due to the inclusion of AGN feedback in the hydro simulations (with sub grid physics). \cite{Osato2015} found that baryon physics can induce significant biases when applied to parameter constraints, and \cite{Fong2019} state that these biases are still present even with baryon physics, unless massive neutrinos are also considered. So a complete understanding of the impact baryons may have on voids identified in WL maps will be important before cosmological constraints can be made. Further more, it is possible that baryons may have different impacts on the peak and void statistics, which is motivation for studying the use of weak lensing voids as a complimentary probe to WL peaks.

It will also be interesting to consider other MG or dark energy theories such as those with different screening mechanisms. The models tested in this paper employ the Vainshtein screening mechanism which depends on derivatives of the scalar field, where other screening mechanism such as chameleon screening in $f(R)$ gravity may leave different imprints on the WL convergence maps, and hence on the statistics of WL voids. For galaxy voids, the tunnel algorithm is a better test of chameleon screening \citep{Cautun2018} than Vainshtein screening \citep{Paillas2019}. So it will be important to consider multiple screening mechanisms, where this method can then be used to place constraints on the screening thresholds for MG theories.

To summarise, the work presented here shows that the study of 2D voids identified in WL maps can be a useful statistic to develop in order to maximise the information that can be gained from future surveys. Further development such as testing multiple screening mechanisms, the impact of baryon physics on the peak and void statistics in MG, and the analysis of potential cosmological parameter constraints will be left for future work.

\section*{Acknowledgements}
We thank Alexandre Barreira for providing the lensing maps used in the analysis of this work. CTD is funded by a Science and Technology Facilities Council (STFC) PhD studentship through grant ST/R504725/1. MC acknowledges support by the European Research Council through an ERC Advanced Investigator Grant, DMIDAS [GA-786910] and by the EU Horizon 2020 research and innovation programme under a Marie Sk{\l}odowska-Curie grant agreement 794474 (DancingGalaxies). BL is supported by an ERC Starting Grant, ERC-StG-PUNCA-716532. MC and BL are additionally supported by the STFC Consolidated Grants [Nos.~ST/I00162X/1, ST/P000541/1]

This work used the DiRAC facility, hosted by Durham University, managed by the Institute for
Computational Cosmology on behalf of the STFC DiRAC HPC Facility (www.dirac.ac.uk). The equipment was funded by BEIS capital funding
via STFC capital grants ST/K00042X/1, ST/P002293/1, ST/R002371/1 and ST/S002502/1, Durham University and STFC operations grant
ST/R000832/1. DiRAC is part of the National e-Infrastructure.





\bibliographystyle{mnras}
\bibliography{mybib}

\begin{thebibliography}{}
\makeatletter
\relax
\def\mn@urlcharsother{\let\do\@makeother \do\$\do\&\do\#\do\^\do\_\do\%\do\~}
\def\mn@doi{\begingroup\mn@urlcharsother \@ifnextchar [ {\mn@doi@}
  {\mn@doi@[]}}
\def\mn@doi@[#1]#2{\def\@tempa{#1}\ifx\@tempa\@empty \href
  {http://dx.doi.org/#2} {doi:#2}\else \href {http://dx.doi.org/#2} {#1}\fi
  \endgroup}
\def\mn@eprint#1#2{\mn@eprint@#1:#2::\@nil}
\def\mn@eprint@arXiv#1{\href {http://arxiv.org/abs/#1} {{\tt arXiv:#1}}}
\def\mn@eprint@dblp#1{\href {http://dblp.uni-trier.de/rec/bibtex/#1.xml}
  {dblp:#1}}
\def\mn@eprint@#1:#2:#3:#4\@nil{\def\@tempa {#1}\def\@tempb {#2}\def\@tempc
  {#3}\ifx \@tempc \@empty \let \@tempc \@tempb \let \@tempb \@tempa \fi \ifx
  \@tempb \@empty \def\@tempb {arXiv}\fi \@ifundefined
  {mn@eprint@\@tempb}{\@tempb:\@tempc}{\expandafter \expandafter \csname
  mn@eprint@\@tempb\endcsname \expandafter{\@tempc}}}

\bibitem[\protect\citeauthoryear{Achitouv}{Achitouv}{2016}]{Achitouv2016}
Achitouv I.,  2016, \mn@doi [PRD] {10.1103/PhysRevD.94.103524}, 94, 103524

\bibitem[\protect\citeauthoryear{{Albrecht} et~al.,}{{Albrecht}
  et~al.}{2006}]{Albrecht2006}
{Albrecht} A.,  et~al., 2006, preprint, \href
  {https://ui.adsabs.harvard.edu/\#abs/2006astro.ph..9591A} {} (\mn@eprint
  {arXiv} {astro-ph/0609591})

\bibitem[\protect\citeauthoryear{Amendola et~al.,}{Amendola
  et~al.}{2013}]{Amendola2013}
Amendola L.,  et~al., 2013, \mn@doi [LRV] {10.12942/lrr-2013-6}, 16, 6

\bibitem[\protect\citeauthoryear{Anderson}{Anderson}{2003}]{Anderson2003}
Anderson T.~W.,  2003, An introduction to multivariate statistical analysis.
Wiley-Interscience

\bibitem[\protect\citeauthoryear{{Bacon}, {Refregier}  \& {Ellis}}{{Bacon}
  et~al.}{2000}]{Bacon2000}
{Bacon} D.~J.,  {Refregier} A.~R.,   {Ellis} R.~S.,  2000, \mn@doi [\mnras]
  {10.1046/j.1365-8711.2000.03851.x}, 318, 625

\bibitem[\protect\citeauthoryear{Banerjee \& Dalal}{Banerjee \&
  Dalal}{2016}]{Banerjee2016}
Banerjee A.,  Dalal N.,  2016, \mn@doi [JCAP] {10.1088/1475-7516/2016/11/015},
  2016, 015

\bibitem[\protect\citeauthoryear{Barreira, Li, Baugh  \& Pascoli}{Barreira
  et~al.}{2014}]{Barreira2014}
Barreira A.,  Li B.,  Baugh C.~M.,   Pascoli S.,  2014, \mn@doi [PRD]
  {10.1103/PhysRevD.90.023528}, 90, 023528

\bibitem[\protect\citeauthoryear{Barreira, Cautun, Li, Baugh  \&
  Pascoli}{Barreira et~al.}{2015}]{Barreira2015}
Barreira A.,  Cautun M.,  Li B.,  Baugh C.~M.,   Pascoli S.,  2015, \mn@doi
  [JCAP] {10.1088/1475-7516/2015/08/028}, 2015, 028

\bibitem[\protect\citeauthoryear{{Barreira}, {Llinares}, {Bose}  \&
  {Li}}{{Barreira} et~al.}{2016}]{Barreira2016}
{Barreira} A.,  {Llinares} C.,  {Bose} S.,   {Li} B.,  2016, \mn@doi [JCAP]
  {10.1088/1475-7516/2016/05/001}, \href
  {https://ui.adsabs.harvard.edu/abs/2016JCAP...05..001B} {2016, 001}

\bibitem[\protect\citeauthoryear{{Barreira}, {Bose}, {Li}  \&
  {Llinares}}{{Barreira} et~al.}{2017}]{Barreira2017}
{Barreira} A.,  {Bose} S.,  {Li} B.,   {Llinares} C.,  2017, \mn@doi [JCAP]
  {10.1088/1475-7516/2017/02/031}, \href
  {https://ui.adsabs.harvard.edu/abs/2017JCAP...02..031B} {2017, 031}

\bibitem[\protect\citeauthoryear{{Bartelmann} \& {Schneider}}{{Bartelmann} \&
  {Schneider}}{2001}]{Bartelmann2001}
{Bartelmann} M.,  {Schneider} P.,  2001, \mn@doi [PR]
  {10.1016/S0370-1573(00)00082-X}, \href
  {https://ui.adsabs.harvard.edu/\#abs/2001PhR...340..291B} {340, 291}

\bibitem[\protect\citeauthoryear{{Bertotti}, {Iess}  \& {Tortora}}{{Bertotti}
  et~al.}{2003}]{Bertotti2003}
{Bertotti} B.,  {Iess} L.,   {Tortora} P.,  2003, \mn@doi [\nat]
  {10.1038/nature01997}, \href
  {http://adsabs.harvard.edu/abs/2003Natur.425..374B} {425, 374}

\bibitem[\protect\citeauthoryear{{Bond}, {Kofman}  \& {Pogosyan}}{{Bond}
  et~al.}{1996}]{Bond1996}
{Bond} J.~R.,  {Kofman} L.,   {Pogosyan} D.,  1996, \mn@doi [\nat]
  {10.1038/380603a0}, \href {http://adsabs.harvard.edu/abs/1996Natur.380..603B}
  {380, 603}

\bibitem[\protect\citeauthoryear{Bos, van~de Weygaert, Dolag  \& Pettorino}{Bos
  et~al.}{2012}]{Bos2012}
Bos E. G.~P.,  van~de Weygaert R.,  Dolag K.,   Pettorino V.,  2012, \mn@doi
  [\mnras] {10.1111/j.1365-2966.2012.21478.x}, 426, 440

\bibitem[\protect\citeauthoryear{Brax}{Brax}{2013}]{Brax2013}
Brax P.,  2013, \mn@doi [CQG] {10.1088/0264-9381/30/21/214005}, 30, 214005

\bibitem[\protect\citeauthoryear{Cai, Padilla  \& Li}{Cai
  et~al.}{2015}]{Cai2015}
Cai Y.-C.,  Padilla N.,   Li B.,  2015, \mn@doi [\mnras]
  {10.1093/mnras/stv777}, 451, 1036

\bibitem[\protect\citeauthoryear{{Caldwell} \& {Kamionkowski}}{{Caldwell} \&
  {Kamionkowski}}{2009}]{Caldwell2009}
{Caldwell} R.~R.,  {Kamionkowski} M.,  2009, \mn@doi [ARNPS]
  {10.1146/annurev-nucl-010709-151330}, \href
  {https://ui.adsabs.harvard.edu/abs/2009ARNPS..59..397C} {59, 397}

\bibitem[\protect\citeauthoryear{{Cardone}, {Camera}, {Mainini}, {Romano},
  {Diaferio}, {Maoli}  \& {Scaramella}}{{Cardone} et~al.}{2013}]{Cardone2013}
{Cardone} V.~F.,  {Camera} S.,  {Mainini} R.,  {Romano} A.,  {Diaferio} A.,
  {Maoli} R.,   {Scaramella} R.,  2013, \mn@doi [\mnras]
  {10.1093/mnras/stt084}, \href
  {https://ui.adsabs.harvard.edu/\#abs/2013MNRAS.430.2896C} {430, 2896}

\bibitem[\protect\citeauthoryear{{Cautun}, {van de Weygaert}  \&
  {Jones}}{{Cautun} et~al.}{2013}]{Cautun2013}
{Cautun} M.,  {van de Weygaert} R.,   {Jones} B. J.~T.,  2013, \mn@doi [\mnras]
  {10.1093/mnras/sts416}, \href
  {https://ui.adsabs.harvard.edu/abs/2013MNRAS.429.1286C} {429, 1286}

\bibitem[\protect\citeauthoryear{Cautun, van~de Weygaert, Jones  \&
  Frenk}{Cautun et~al.}{2014}]{Cautun2014}
Cautun M.,  van~de Weygaert R.,  Jones B. J.~T.,   Frenk C.~S.,  2014, \mn@doi
  [\mnras] {10.1093/mnras/stu768}, 441, 2923

\bibitem[\protect\citeauthoryear{{Cautun}, {Cai}  \& {Frenk}}{{Cautun}
  et~al.}{2016}]{Cautun2016}
{Cautun} M.,  {Cai} Y.-C.,   {Frenk} C.~S.,  2016, \mn@doi [\mnras]
  {10.1093/mnras/stw154}, \href
  {https://ui.adsabs.harvard.edu/abs/2016MNRAS.457.2540C} {457, 2540}

\bibitem[\protect\citeauthoryear{{Cautun}, {Paillas}, {Cai}, {Bose}, {Armijo},
  {Li}  \& {Padilla}}{{Cautun} et~al.}{2018}]{Cautun2018}
{Cautun} M.,  {Paillas} E.,  {Cai} Y.-C.,  {Bose} S.,  {Armijo} J.,  {Li} B.,
  {Padilla} N.,  2018, \mn@doi [\mnras] {10.1093/mnras/sty463}, \href
  {http://adsabs.harvard.edu/abs/2018MNRAS.476.3195C} {476, 3195}

\bibitem[\protect\citeauthoryear{Clampitt, Cai  \& Li}{Clampitt
  et~al.}{2013}]{Clampitt2013}
Clampitt J.,  Cai Y.-C.,   Li B.,  2013, \mn@doi [\mnras]
  {10.1093/mnras/stt219}, 431, 749

\bibitem[\protect\citeauthoryear{{Correa}, {Paz}, {Padilla}, {Ruiz}, {Angulo}
  \& {S{\'a}nchez}}{{Correa} et~al.}{2019}]{Correa2019}
{Correa} C.~M.,  {Paz} D.~J.,  {Padilla} N.~D.,  {Ruiz} A.~N.,  {Angulo} R.~E.,
    {S{\'a}nchez} A.~G.,  2019, \mn@doi [\mnras] {10.1093/mnras/stz821}, \href
  {https://ui.adsabs.harvard.edu/abs/2019MNRAS.485.5761C} {485, 5761}

\bibitem[\protect\citeauthoryear{{Davies}, {Cautun}  \& {Li}}{{Davies}
  et~al.}{2018}]{Davies2018}
{Davies} C.~T.,  {Cautun} M.,   {Li} B.,  2018, \mn@doi [\mnras]
  {10.1093/mnrasl/sly135}, \href
  {https://ui.adsabs.harvard.edu/\#abs/2018MNRAS.480L.101D} {480, L101}

\bibitem[\protect\citeauthoryear{{Davies}, {Cautun}  \& {Li}}{{Davies}
  et~al.}{2019}]{Davies2019}
{Davies} C.~T.,  {Cautun} M.,   {Li} B.,  2019, preprint, \href
  {https://ui.adsabs.harvard.edu/abs/2019arXiv190501710D} {} (\mn@eprint
  {arXiv} {1905.01710})

\bibitem[\protect\citeauthoryear{{Davis}, {Efstathiou}, {Frenk}  \&
  {White}}{{Davis} et~al.}{1985}]{Davis1985}
{Davis} M.,  {Efstathiou} G.,  {Frenk} C.~S.,   {White} S.~D.~M.,  1985,
  \mn@doi [\apj] {10.1086/163168}, \href
  {http://adsabs.harvard.edu/abs/1985ApJ...292..371D} {292, 371}

\bibitem[\protect\citeauthoryear{Demchenko, Cai, Heymans  \& Peacock}{Demchenko
  et~al.}{2016}]{Demchenko2016}
Demchenko V.,  Cai Y.-C.,  Heymans C.,   Peacock J.~A.,  2016, \mn@doi [\mnras]
  {10.1093/mnras/stw2030}, 463, 512

\bibitem[\protect\citeauthoryear{{Dvali}, {Gabadadze}  \& {Porrati}}{{Dvali}
  et~al.}{2000}]{Dvali2000}
{Dvali} G.,  {Gabadadze} G.,   {Porrati} M.,  2000, \mn@doi [PLB]
  {10.1016/S0370-2693(00)00631-6}, \href
  {https://ui.adsabs.harvard.edu/abs/2000PhLB..484..112D} {484, 112}

\bibitem[\protect\citeauthoryear{Erben et~al.,}{Erben et~al.}{2013}]{Erben2013}
Erben T.,  et~al., 2013, \mn@doi [MNRAS] {10.1093/mnras/stt928}, 433, 2545

\bibitem[\protect\citeauthoryear{{Falck}, {Koyama}, {Zhao}  \&
  {Cautun}}{{Falck} et~al.}{2018}]{Falck2018}
{Falck} B.,  {Koyama} K.,  {Zhao} G.-B.,   {Cautun} M.,  2018, \mn@doi [\mnras]
  {10.1093/mnras/stx3288}, \href
  {https://ui.adsabs.harvard.edu/abs/2018MNRAS.475.3262F} {475, 3262}

\bibitem[\protect\citeauthoryear{{Fong}, {Choi}, {Catlett}, {Lee}, {Peel},
  {Bowyer}, {King}  \& {McCarthy}}{{Fong} et~al.}{2019}]{Fong2019}
{Fong} M.,  {Choi} M.,  {Catlett} V.,  {Lee} B.,  {Peel} A.,  {Bowyer} R.,
  {King} L.~J.,   {McCarthy} I.~G.,  2019, preprint, \href
  {https://ui.adsabs.harvard.edu/abs/2019arXiv190703161F} {} (\mn@eprint
  {arXiv} {1907.03161})

\bibitem[\protect\citeauthoryear{{Fu} et~al.,}{{Fu} et~al.}{2008}]{Fu2008}
{Fu} L.,  et~al., 2008, \mn@doi [\aap] {10.1051/0004-6361:20078522}, \href
  {https://ui.adsabs.harvard.edu/\#abs/2008A&A...479....9F} {479, 9}

\bibitem[\protect\citeauthoryear{{Giocoli}, {Moscardini}, {Baldi}, {Meneghetti}
   \& {Metcalf}}{{Giocoli} et~al.}{2018}]{Giocoli2018}
{Giocoli} C.,  {Moscardini} L.,  {Baldi} M.,  {Meneghetti} M.,   {Metcalf}
  R.~B.,  2018, \mn@doi [\mnras] {10.1093/mnras/sty1312}, \href
  {https://ui.adsabs.harvard.edu/\#abs/2018MNRAS.478.5436G} {478, 5436}

\bibitem[\protect\citeauthoryear{Haider, Steinhauser, Vogelsberger, Genel,
  Springel, Torrey  \& Hernquist}{Haider et~al.}{2016}]{Haider2016}
Haider M.,  Steinhauser D.,  Vogelsberger M.,  Genel S.,  Springel V.,  Torrey
  P.,   Hernquist L.,  2016, \mn@doi [\mnras] {10.1093/mnras/stw077}, 457, 3024

\bibitem[\protect\citeauthoryear{Hamana, Oguri, Shirasaki  \& Sato}{Hamana
  et~al.}{2012}]{Hamana2012}
Hamana T.,  Oguri M.,  Shirasaki M.,   Sato M.,  2012, \mn@doi [\mnras]
  {10.1111/j.1365-2966.2012.21582.x}, 425, 2287

\bibitem[\protect\citeauthoryear{{Hamana}, {Sakurai}, {Koike}  \&
  {Miller}}{{Hamana} et~al.}{2015}]{Hamana2015}
{Hamana} T.,  {Sakurai} J.,  {Koike} M.,   {Miller} L.,  2015, \mn@doi [\pasj]
  {10.1093/pasj/psv034}, \href {http://ads.nao.ac.jp/abs/2015PASJ...67...34H}
  {67, 34}

\bibitem[\protect\citeauthoryear{{Hamaus}, {Sutter}, {Lavaux}  \& {Wand
  elt}}{{Hamaus} et~al.}{2015}]{Hamaus2015}
{Hamaus} N.,  {Sutter} P.~M.,  {Lavaux} G.,   {Wand elt} B.~D.,  2015, \mn@doi
  [\jcap] {10.1088/1475-7516/2015/11/036}, \href
  {https://ui.adsabs.harvard.edu/abs/2015JCAP...11..036H} {2015, 036}

\bibitem[\protect\citeauthoryear{{Hamaus}, {Pisani}, {Sutter}, {Lavaux},
  {Escoffier}, {Wand elt}  \& {Weller}}{{Hamaus} et~al.}{2016}]{Hamaus2016}
{Hamaus} N.,  {Pisani} A.,  {Sutter} P.~M.,  {Lavaux} G.,  {Escoffier} S.,
  {Wand elt} B.~D.,   {Weller} J.,  2016, \mn@doi [\prl]
  {10.1103/PhysRevLett.117.091302}, \href
  {https://ui.adsabs.harvard.edu/abs/2016PhRvL.117i1302H} {117, 091302}

\bibitem[\protect\citeauthoryear{{Hartlap}, {Simon}  \& {Schneider}}{{Hartlap}
  et~al.}{2007}]{Hartlap2007}
{Hartlap} J.,  {Simon} P.,   {Schneider} P.,  2007, \mn@doi [\aap]
  {10.1051/0004-6361:20066170}, \href
  {https://ui.adsabs.harvard.edu/abs/2007A&A...464..399H} {464, 399}

\bibitem[\protect\citeauthoryear{Heymans et~al.,}{Heymans
  et~al.}{2012}]{Heymans2012}
Heymans C.,  et~al., 2012, \mn@doi [\mnras] {10.1111/j.1365-2966.2012.21952.x},
  427, 146

\bibitem[\protect\citeauthoryear{{Higuchi} \& {Shirasaki}}{{Higuchi} \&
  {Shirasaki}}{2016}]{Higuchi2016}
{Higuchi} Y.,  {Shirasaki} M.,  2016, \mn@doi [\mnras] {10.1093/mnras/stw814},
  \href {https://ui.adsabs.harvard.edu/\#abs/2016MNRAS.459.2762H} {459, 2762}

\bibitem[\protect\citeauthoryear{{Hildebrandt} et~al.,}{{Hildebrandt}
  et~al.}{2017}]{Hildebrandt2017}
{Hildebrandt} H.,  et~al., 2017, \mn@doi [\mnras] {10.1093/mnras/stw2805},
  \href {http://adsabs.harvard.edu/abs/2017MNRAS.465.1454H} {465, 1454}

\bibitem[\protect\citeauthoryear{Hoekstra et~al.,}{Hoekstra
  et~al.}{2006}]{Hoekstra2006}
Hoekstra H.,  et~al., 2006, \mn@doi [\apj] {10.1086/503249}, 647, 116

\bibitem[\protect\citeauthoryear{{Kaiser}, {Wilson}  \& {Luppino}}{{Kaiser}
  et~al.}{2000}]{Kaiser2000}
{Kaiser} N.,  {Wilson} G.,   {Luppino} G.~A.,  2000, preprint, \href
  {https://ui.adsabs.harvard.edu/\#abs/2000astro.ph..3338K} {} (\mn@eprint
  {arXiv} {astro-ph/0003338})

\bibitem[\protect\citeauthoryear{{Kilbinger} et~al.,}{{Kilbinger}
  et~al.}{2013}]{Kilbinger2013}
{Kilbinger} M.,  et~al., 2013, \mn@doi [\mnras] {10.1093/mnras/stt041}, \href
  {https://ui.adsabs.harvard.edu/\#abs/2013MNRAS.430.2200K} {430, 2200}

\bibitem[\protect\citeauthoryear{{Kirshner}, {Oemler}, {Schechter}  \&
  {Shectman}}{{Kirshner} et~al.}{1981}]{Kirshner1981}
{Kirshner} R.~P.,  {Oemler} Jr. A.,  {Schechter} P.~L.,   {Shectman} S.~A.,
  1981, \mn@doi [\apjl] {10.1086/183623}, \href
  {http://adsabs.harvard.edu/abs/1981ApJ...248L..57K} {248, L57}

\bibitem[\protect\citeauthoryear{{Kreisch}, {Pisani}, {Carbone}, {Liu},
  {Hawken}, {Massara}, {Spergel}  \& {Wand elt}}{{Kreisch}
  et~al.}{2018}]{Kreisch2018}
{Kreisch} C.~D.,  {Pisani} A.,  {Carbone} C.,  {Liu} J.,  {Hawken} A.~J.,
  {Massara} E.,  {Spergel} D.~N.,   {Wand elt} B.~D.,  2018, preprint, \href
  {https://ui.adsabs.harvard.edu/abs/2018arXiv180807464K} {} (\mn@eprint
  {arXiv} {1808.07464})

\bibitem[\protect\citeauthoryear{{LSST Dark Energy Science
  Collaboration}}{{LSST Dark Energy Science Collaboration}}{2012}]{LSST2012}
{LSST Dark Energy Science Collaboration} 2012, preprint, \href
  {https://ui.adsabs.harvard.edu/\#abs/2012arXiv1211.0310L} {} (\mn@eprint
  {arXiv} {1211.0310})

\bibitem[\protect\citeauthoryear{{LSST Science Collaboration} et~al.,}{{LSST
  Science Collaboration} et~al.}{2009}]{LSST2009}
{LSST Science Collaboration} et~al., 2009, preprint, \href
  {https://ui.adsabs.harvard.edu/\#abs/2009arXiv0912.0201L} {} (\mn@eprint
  {arXiv} {0912.0201})

\bibitem[\protect\citeauthoryear{{Lavaux} \& {Wandelt}}{{Lavaux} \&
  {Wandelt}}{2012}]{Lavaux2012}
{Lavaux} G.,  {Wandelt} B.~D.,  2012, \mn@doi [\apj]
  {10.1088/0004-637X/754/2/109}, \href
  {https://ui.adsabs.harvard.edu/abs/2012ApJ...754..109L} {754, 109}

\bibitem[\protect\citeauthoryear{Li}{Li}{2011}]{Li2011}
Li B.,  2011, \mn@doi [\mnras] {10.1111/j.1365-2966.2010.17867.x}, 411, 2615

\bibitem[\protect\citeauthoryear{{Li}, {King}, {Zhao}  \& {Zhao}}{{Li}
  et~al.}{2011}]{Li-ray-tracing}
{Li} B.,  {King} L.~J.,  {Zhao} G.-B.,   {Zhao} H.,  2011, \mn@doi [MNRAS]
  {10.1111/j.1365-2966.2011.18754.x}, 415, 881

\bibitem[\protect\citeauthoryear{{Li}, {Zhao}, {Teyssier}  \& {Koyama}}{{Li}
  et~al.}{2012}]{Li2012}
{Li} B.,  {Zhao} G.-B.,  {Teyssier} R.,   {Koyama} K.,  2012, \mn@doi [JCAP]
  {10.1088/1475-7516/2012/01/051}, \href
  {https://ui.adsabs.harvard.edu/abs/2012JCAP...01..051L} {2012, 051}

\bibitem[\protect\citeauthoryear{{Li}, {Zhao}  \& {Koyama}}{{Li}
  et~al.}{2013}]{Li2013}
{Li} B.,  {Zhao} G.-B.,   {Koyama} K.,  2013, \mn@doi [JCAP]
  {10.1088/1475-7516/2013/05/023}, \href
  {https://ui.adsabs.harvard.edu/abs/2013JCAP...05..023L} {2013, 023}

\bibitem[\protect\citeauthoryear{{Li}, {Liu}, {Zorrilla Matilla}  \&
  {Coulton}}{{Li} et~al.}{2018}]{Li2018}
{Li} Z.,  {Liu} J.,  {Zorrilla Matilla} J.~M.,   {Coulton} W.~R.,  2018,
  preprint, \href {https://ui.adsabs.harvard.edu/\#abs/2018arXiv181001781L} {}
  (\mn@eprint {arXiv} {1810.01781})

\bibitem[\protect\citeauthoryear{{Liu}, {Petri}, {Haiman}, {Hui}, {Kratochvil}
  \& {May}}{{Liu} et~al.}{2015a}]{J.Liu2015}
{Liu} J.,  {Petri} A.,  {Haiman} Z.,  {Hui} L.,  {Kratochvil} J.~M.,   {May}
  M.,  2015a, \mn@doi [\prd] {10.1103/PhysRevD.91.063507}, \href
  {https://ui.adsabs.harvard.edu/\#abs/2015PhRvD..91f3507L} {91, 063507}

\bibitem[\protect\citeauthoryear{Liu et~al.,}{Liu et~al.}{2015b}]{X.Liu2015}
Liu X.,  et~al., 2015b, \mn@doi [\mnras] {10.1093/mnras/stv784}, \href
  {https://ui.adsabs.harvard.edu/\#abs/2015MNRAS.450.2888L} {450, 2888}

\bibitem[\protect\citeauthoryear{{Liu}, {Hill}, {Sherwin}, {Petri}, {B{\"o}hm}
  \& {Haiman}}{{Liu} et~al.}{2016a}]{J.Liu2016b}
{Liu} J.,  {Hill} J.~C.,  {Sherwin} B.~D.,  {Petri} A.,  {B{\"o}hm} V.,
  {Haiman} Z.,  2016a, \mn@doi [PRD] {10.1103/PhysRevD.94.103501}, \href
  {https://ui.adsabs.harvard.edu/abs/2016PhRvD..94j3501L} {94, 103501}

\bibitem[\protect\citeauthoryear{{Liu} et~al.,}{{Liu}
  et~al.}{2016b}]{X.Liu2016}
{Liu} X.,  et~al., 2016b, \mn@doi [PRL] {10.1103/PhysRevLett.117.051101}, \href
  {https://ui.adsabs.harvard.edu/\#abs/2016PhRvL.117e1101L} {117, 051101}

\bibitem[\protect\citeauthoryear{Massara, Villaescusa-Navarro, Viel  \&
  Sutter}{Massara et~al.}{2015}]{Massara2015}
Massara E.,  Villaescusa-Navarro F.,  Viel M.,   Sutter P.,  2015, \mn@doi
  [JCAP] {10.1088/1475-7516/2015/11/018}, 2015, 018

\bibitem[\protect\citeauthoryear{{Nadathur}, {Carter}, {Percival}, {Winther}
  \& {Bautista}}{{Nadathur} et~al.}{2019}]{Nadathur2019}
{Nadathur} S.,  {Carter} P.~M.,  {Percival} W.~J.,  {Winther} H.~A.,
  {Bautista} J.,  2019, preprint, \href
  {https://ui.adsabs.harvard.edu/abs/2019arXiv190401030N} {} (\mn@eprint
  {arXiv} {1904.01030})

\bibitem[\protect\citeauthoryear{{Osato}, {Shirasaki}  \& {Yoshida}}{{Osato}
  et~al.}{2015}]{Osato2015}
{Osato} K.,  {Shirasaki} M.,   {Yoshida} N.,  2015, \mn@doi [\apj]
  {10.1088/0004-637X/806/2/186}, \href
  {https://ui.adsabs.harvard.edu/abs/2015ApJ...806..186O} {806, 186}

\bibitem[\protect\citeauthoryear{Padilla, Ceccarelli  \& Lambas}{Padilla
  et~al.}{2005}]{Padilla2005}
Padilla N.~D.,  Ceccarelli L.,   Lambas D.~G.,  2005, \mn@doi [\mnras]
  {10.1111/j.1365-2966.2005.09500.x}, 363, 977

\bibitem[\protect\citeauthoryear{{Paillas}, {Cautun}, {Li}, {Cai}, {Padilla},
  {Armijo}  \& {Bose}}{{Paillas} et~al.}{2019}]{Paillas2019}
{Paillas} E.,  {Cautun} M.,  {Li} B.,  {Cai} Y.-C.,  {Padilla} N.,  {Armijo}
  J.,   {Bose} S.,  2019, \mn@doi [\mnras] {10.1093/mnras/stz022}, \href
  {https://ui.adsabs.harvard.edu/abs/2019MNRAS.484.1149P} {484, 1149}

\bibitem[\protect\citeauthoryear{{Peel}, {Pettorino}, {Giocoli}, {Starck}  \&
  {Baldi}}{{Peel} et~al.}{2018}]{Peel2018}
{Peel} A.,  {Pettorino} V.,  {Giocoli} C.,  {Starck} J.-L.,   {Baldi} M.,
  2018, \mn@doi [\aap] {10.1051/0004-6361/201833481}, \href
  {https://ui.adsabs.harvard.edu/\#abs/2018A&A...619A..38P} {619, A38}

\bibitem[\protect\citeauthoryear{{Perlmutter} et~al.,}{{Perlmutter}
  et~al.}{1999}]{Perlmutter1999}
{Perlmutter} S.,  et~al., 1999, \mn@doi [\apj] {10.1086/307221}, \href
  {https://ui.adsabs.harvard.edu/abs/1999ApJ...517..565P} {517, 565}

\bibitem[\protect\citeauthoryear{Pisani, Sutter, Hamaus, Alizadeh, Biswas,
  Wandelt  \& Hirata}{Pisani et~al.}{2015}]{Pisani2015}
Pisani A.,  Sutter P.~M.,  Hamaus N.,  Alizadeh E.,  Biswas R.,  Wandelt B.~D.,
    Hirata C.~M.,  2015, \mn@doi [Phys. Rev. D] {10.1103/PhysRevD.92.083531},
  92, 083531

\bibitem[\protect\citeauthoryear{{Platen}, {van de Weygaert}  \&
  {Jones}}{{Platen} et~al.}{2007}]{Platen2007}
{Platen} E.,  {van de Weygaert} R.,   {Jones} B. J.~T.,  2007, \mn@doi [\mnras]
  {10.1111/j.1365-2966.2007.12125.x}, \href
  {https://ui.adsabs.harvard.edu/abs/2007MNRAS.380..551P} {380, 551}

\bibitem[\protect\citeauthoryear{{Refregier}, {Amara}, {Kitching}, {Rassat},
  {Scaramella}, {Weller}  \& {Euclid Imaging Consortium}}{{Refregier}
  et~al.}{2010}]{Refregier2010}
{Refregier} A.,  {Amara} A.,  {Kitching} T.~D.,  {Rassat} A.,  {Scaramella} R.,
   {Weller} J.,   {Euclid Imaging Consortium} f.~t.,  2010, preprint, \href
  {https://ui.adsabs.harvard.edu/\#abs/2010arXiv1001.0061R} {} (\mn@eprint
  {arXiv} {1001.0061})

\bibitem[\protect\citeauthoryear{{Riess} et~al.,}{{Riess}
  et~al.}{1998}]{Riess1998}
{Riess} A.~G.,  et~al., 1998, \mn@doi [\aj] {10.1086/300499}, \href
  {https://ui.adsabs.harvard.edu/abs/1998AJ....116.1009R} {116, 1009}

\bibitem[\protect\citeauthoryear{{Schneider}}{{Schneider}}{1996}]{Schneider1996}
{Schneider} P.,  1996, \mn@doi [\mnras] {10.1093/mnras/283.3.837}, \href
  {https://ui.adsabs.harvard.edu/abs/1996MNRAS.283..837S} {283, 837}

\bibitem[\protect\citeauthoryear{{Schneider}, {van Waerbeke}, {Kilbinger}  \&
  {Mellier}}{{Schneider} et~al.}{2002}]{Schneider2002}
{Schneider} P.,  {van Waerbeke} L.,  {Kilbinger} M.,   {Mellier} Y.,  2002,
  \mn@doi [\aap] {10.1051/0004-6361:20021341}, \href
  {https://ui.adsabs.harvard.edu/\#abs/2002A&A...396....1S} {396, 1}

\bibitem[\protect\citeauthoryear{{Semboloni} et~al.,}{{Semboloni}
  et~al.}{2006}]{Semboloni2006}
{Semboloni} E.,  et~al., 2006, \mn@doi [\aap] {10.1051/0004-6361:20054479},
  \href {https://ui.adsabs.harvard.edu/\#abs/2006A&A...452...51S} {452, 51}

\bibitem[\protect\citeauthoryear{Shan et~al.,}{Shan et~al.}{2012}]{Shan2012}
Shan H.,  et~al., 2012, \mn@doi [ApJ] {10.1088/0004-637x/748/1/56}, 748, 56

\bibitem[\protect\citeauthoryear{Shan et~al.,}{Shan et~al.}{2014}]{Shan2014}
Shan H.,  et~al., 2014, \mn@doi [MNRAS] {10.1093/mnras/stu1040}, 442, 2534

\bibitem[\protect\citeauthoryear{{Shirasaki}}{{Shirasaki}}{2017}]{Shirasaki2017}
{Shirasaki} M.,  2017, \mn@doi [\mnras] {10.1093/mnras/stw2950}, \href
  {https://ui.adsabs.harvard.edu/\#abs/2017MNRAS.465.1974S} {465, 1974}

\bibitem[\protect\citeauthoryear{{Shirasaki}, {Yoshida}  \&
  {Ikeda}}{{Shirasaki} et~al.}{2018}]{Shirasaki2018}
{Shirasaki} M.,  {Yoshida} N.,   {Ikeda} S.,  2018, preprint, \href
  {https://ui.adsabs.harvard.edu/\#abs/2018arXiv181205781S} {} (\mn@eprint
  {arXiv} {1812.05781})

\bibitem[\protect\citeauthoryear{{Takahashi}, {Hamana}, {Shirasaki},
  {Namikawa}, {Nishimichi}, {Osato}  \& {Shiroyama}}{{Takahashi}
  et~al.}{2017}]{Takahashi2017}
{Takahashi} R.,  {Hamana} T.,  {Shirasaki} M.,  {Namikawa} T.,  {Nishimichi}
  T.,  {Osato} K.,   {Shiroyama} K.,  2017, \mn@doi [\apj]
  {10.3847/1538-4357/aa943d}, \href
  {http://ads.nao.ac.jp/abs/2017ApJ...850...24T} {850, 24}

\bibitem[\protect\citeauthoryear{Vainshtein}{Vainshtein}{1972}]{Vainshtein1972}
Vainshtein A.,  1972, \mn@doi [PLB]
  {https://doi.org/10.1016/0370-2693(72)90147-5}, 39, 393

\bibitem[\protect\citeauthoryear{{Van Waerbeke}}{{Van
  Waerbeke}}{2000}]{VanWaerbeke2000b}
{Van Waerbeke} L.,  2000, \mn@doi [\mnras] {10.1046/j.1365-8711.2000.03259.x},
  \href {https://ui.adsabs.harvard.edu/abs/2000MNRAS.313..524V} {313, 524}

\bibitem[\protect\citeauthoryear{{Van Waerbeke} et~al.,}{{Van Waerbeke}
  et~al.}{2000}]{VanWaerbeke2000}
{Van Waerbeke} L.,  et~al., 2000, A\&A, \href
  {http://adsabs.harvard.edu/abs/2000A\%26A...358...30V} {358, 30}

\bibitem[\protect\citeauthoryear{Van~Waerbeke et~al.,}{Van~Waerbeke
  et~al.}{2013}]{VanWaerbeke2013}
Van~Waerbeke L.,  et~al., 2013, \mn@doi [MNRAS] {10.1093/mnras/stt971}, 433,
  3373

\bibitem[\protect\citeauthoryear{Villaescusa-Navarro, Vogelsberger, Viel  \&
  Loeb}{Villaescusa-Navarro et~al.}{2013}]{Villaescusa-Navarro2013}
Villaescusa-Navarro F.,  Vogelsberger M.,  Viel M.,   Loeb A.,  2013, \mn@doi
  [\mnras] {10.1093/mnras/stt452}, 431, 3670

\bibitem[\protect\citeauthoryear{{Weinberg}, {Mortonson}, {Eisenstein},
  {Hirata}, {Riess}  \& {Rozo}}{{Weinberg} et~al.}{2013}]{Weinberg2013}
{Weinberg} D.~H.,  {Mortonson} M.~J.,  {Eisenstein} D.~J.,  {Hirata} C.,
  {Riess} A.~G.,   {Rozo} E.,  2013, \mn@doi [PR]
  {10.1016/j.physrep.2013.05.001}, \href
  {http://adsabs.harvard.edu/abs/2013PhR...530...87W} {530, 87}

\bibitem[\protect\citeauthoryear{{Weiss}, {Schneider}, {Sgier}, {Kacprzak},
  {Amara}  \& {Refregier}}{{Weiss} et~al.}{2019}]{Weiss2019}
{Weiss} A.~J.,  {Schneider} A.,  {Sgier} R.,  {Kacprzak} T.,  {Amara} A.,
  {Refregier} A.,  2019, preprint, \href
  {https://ui.adsabs.harvard.edu/abs/2019arXiv190511636W} {} (\mn@eprint
  {arXiv} {1905.11636})

\bibitem[\protect\citeauthoryear{White \& Hu}{White \& Hu}{2000}]{White:1999xa}
White M.~J.,  Hu W.,  2000, \mn@doi [Astrophys. J.] {10.1086/309009}, 537, 1

\bibitem[\protect\citeauthoryear{Will}{Will}{2014}]{Will2014}
Will C.~M.,  2014, \mn@doi [Living Reviews in Relativity]
  {10.12942/lrr-2014-4}, 17, 4

\bibitem[\protect\citeauthoryear{{Wittman}, {Tyson}, {Kirkman}, {Dell'Antonio}
  \& {Bernstein}}{{Wittman} et~al.}{2000}]{Wittman2000}
{Wittman} D.~M.,  {Tyson} J.~A.,  {Kirkman} D.,  {Dell'Antonio} I.,
  {Bernstein} G.,  2000, \mn@doi [\nat] {10.1038/35012001}, \href
  {http://adsabs.harvard.edu/abs/2000Natur.405..143W} {405, 143}

\bibitem[\protect\citeauthoryear{{Yang}, {Kratochvil}, {Huffenberger}, {Haiman}
   \& {May}}{{Yang} et~al.}{2013}]{Yang2013}
{Yang} X.,  {Kratochvil} J.~M.,  {Huffenberger} K.,  {Haiman} Z.,   {May} M.,
  2013, \mn@doi [PRD] {10.1103/PhysRevD.87.023511}, \href
  {https://ui.adsabs.harvard.edu/abs/2013PhRvD..87b3511Y} {87, 023511}

\bibitem[\protect\citeauthoryear{{Zel'dovich}}{{Zel'dovich}}{1970}]{Zeldovich1970}
{Zel'dovich} Y.~B.,  1970, \aap, \href
  {http://adsabs.harvard.edu/abs/1970A%26A.....5...84Z} {5, 84}

\bibitem[\protect\citeauthoryear{Zivick, Sutter, Wandelt, Li  \& Lam}{Zivick
  et~al.}{2015}]{Zivick2015}
Zivick P.,  Sutter P.~M.,  Wandelt B.~D.,  Li B.,   Lam T.~Y.,  2015, \mn@doi
  [\mnras] {10.1093/mnras/stv1209}, 451, 4215

\makeatother
\end{thebibliography}



\appendix

\section{Correlation matrices}\label{sec: cov_mat}
\begin{figure}
    \centering
    \includegraphics[width=\columnwidth]{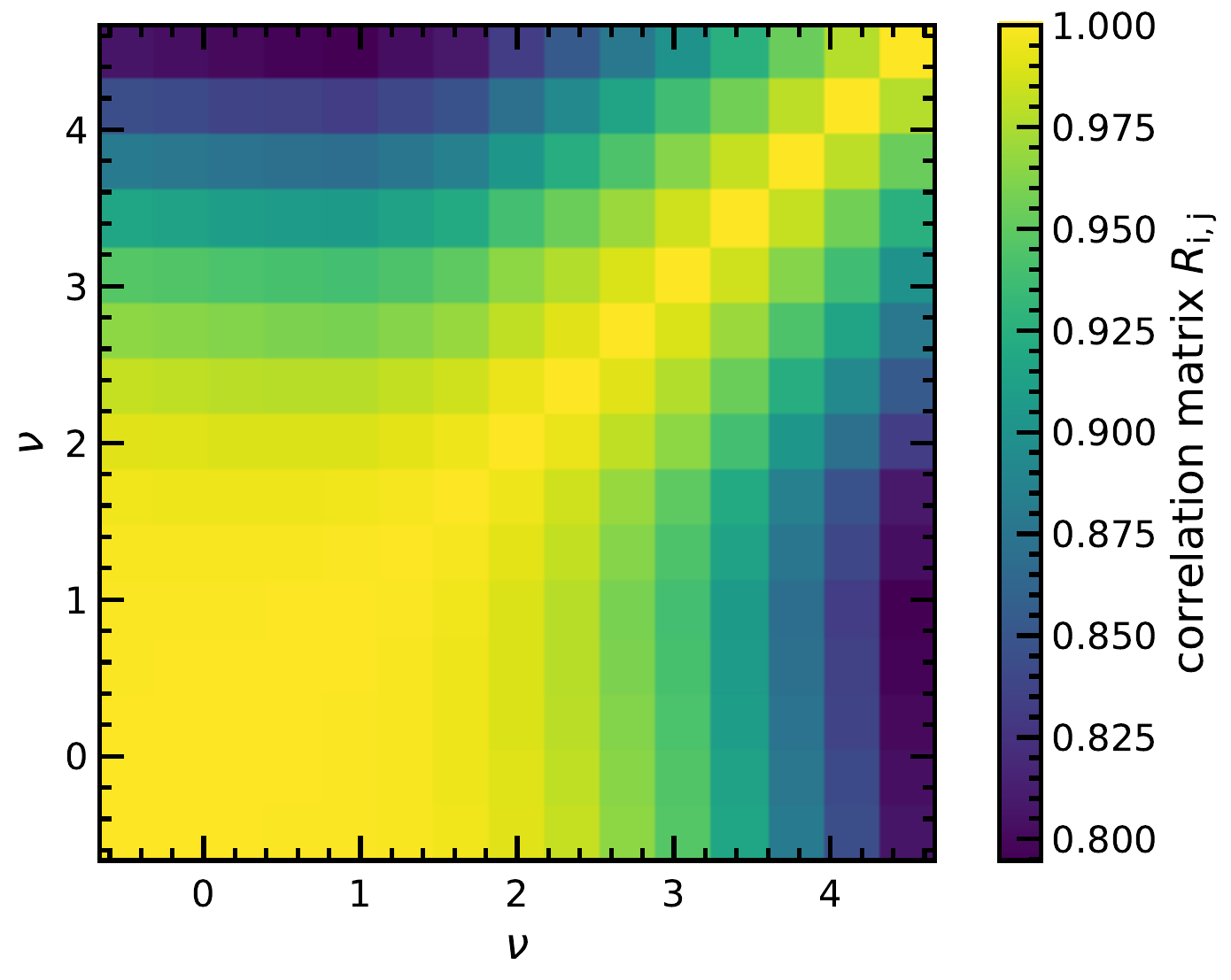}
    \caption{Correlation matrix for the peak abundance extracted from 184 100 deg$^2$ WL maps, where the colour-bar indicates the amplitude of $R_{i,j}$.}
    \label{fig:nu_corr}
\end{figure}

\begin{figure*}
    \centering
    \includegraphics[width=2\columnwidth]{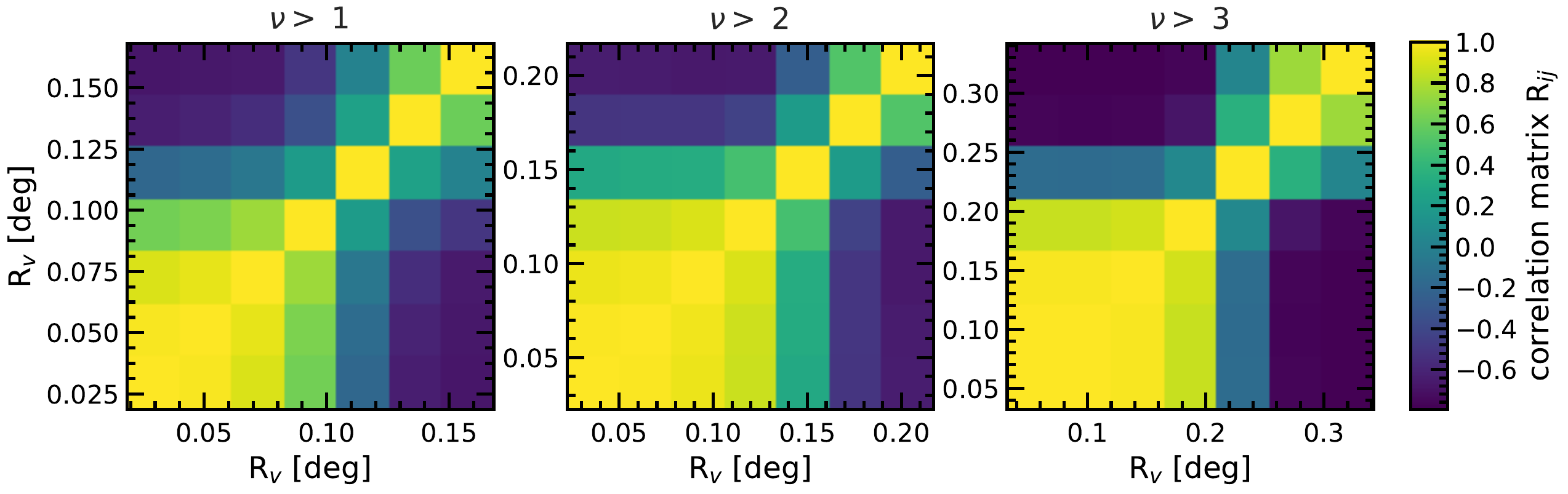}
    \caption{Correlation matrices for the void abundances extracted from 184 100 deg$^2$ WL maps, for three peak height cuts, $\nu > 1$ (left), $\nu > 2$ (middle) and $\nu >3$ (right). The colour-bar indicates the amplitude of $R_{i,j}$.}
    \label{fig:Rv_corr}
\end{figure*}

\begin{figure*}
    \centering
    \includegraphics[width=2\columnwidth]{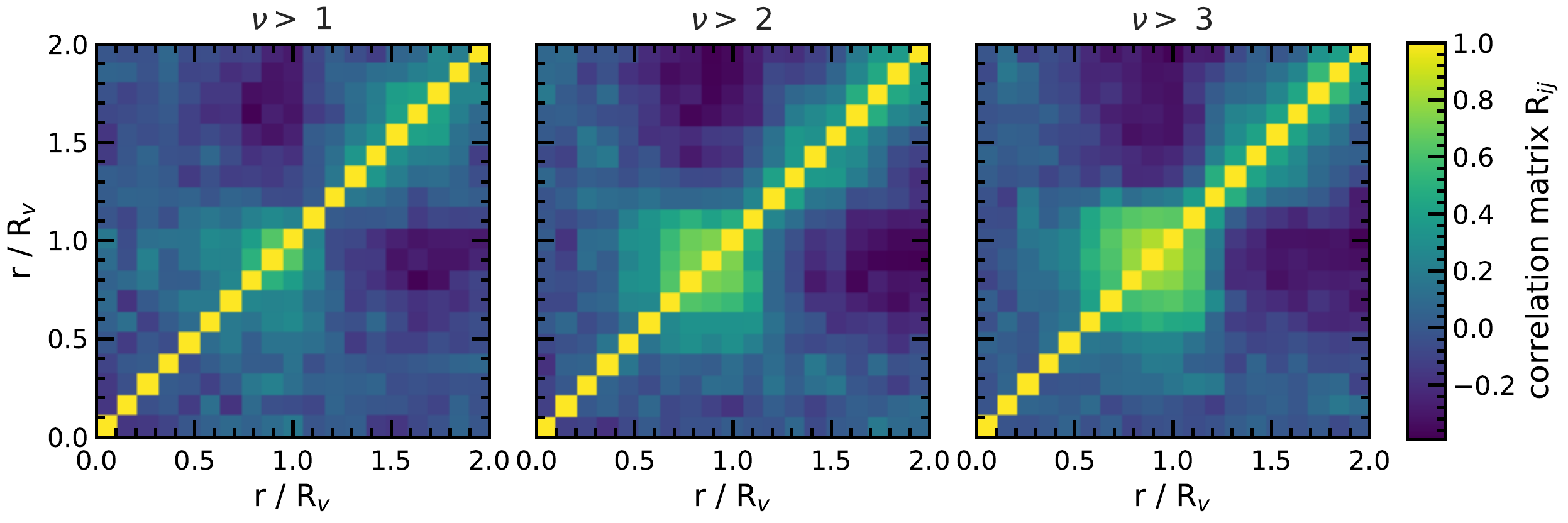}
    \caption{Correlation matrices for the tangential shear profiles extracted from 184 100 deg$^2$ WL maps, for three void catalogues with peak height cuts $\nu > 1$ (left), $\nu > 2$ (middle) and $\nu >3$ (right). The colour-bar indicates the amplitude of $R_{i,j}$.}
    \label{fig:shear_corr}
\end{figure*}

Here we present the covariance matrices for the statistics studied in the paper. All of the covariance matrices are calculated from the statistics extracted from the 184 100 deg$^2$ WL maps from the \citetalias{Takahashi2017} simulations described in section \ref{sec:Numerical simulations}, using Eq. \eqref{eq:SNR}. To aid interpretation we have rescaled all covariance matrices to their corresponding correlation matrix using
\begin{equation}
R_{i,j} = \frac{\rm{cov}(i,j)}{\sigma_i \sigma_j}.
\end{equation}
Figs. \ref{fig:nu_corr}, \ref{fig:Rv_corr}, and \ref{fig:shear_corr} show the correlation matrices for the peak abundances, void abundances, and tangential shear profiles for peaks and voids identified in the \citetalias{Takahashi2017} WL maps.

\bsp	
\label{lastpage}
\end{document}